\documentclass[comment, prd,
nofootinbib, preprintnumbers, superscriptaddress]{revtex4}

\usepackage{amsmath,amssymb,bm,slashed,braket}
\usepackage{graphicx}
\usepackage{subfigure}
\usepackage{epstopdf}
\usepackage{float}
\usepackage{xcolor}
\usepackage{multirow}
\usepackage{dsfont}
\usepackage[colorlinks=true,
            linkcolor=blue,
            urlcolor=blue,
            citecolor=green,          
            bookmarks=true,
            bookmarksnumbered=true,
            breaklinks=true,
            pdfpagemode=Fullscreen,
            pdfstartview=FitBH]{hyperref}
\usepackage[normalem]{ulem}
\allowdisplaybreaks[4]
\usepackage[capitalise]{cleveref}

\newcommand{\vecA}{{\bf A}}

\newcommand{\kvec}{{\bf k}}
\newcommand{\Kvec}{{\bf K}}
\newcommand{\pvec}{{\bf p}}

\newcommand{\rvec}{{\bf r}}

\newcommand{\vvec}{{\bf v}}
\newcommand{\xvec}{{\bf x}}

\newcommand{\vbar}{\bar{v}}

\newcommand{\BG}{\textrm{BG}}
\newcommand{\pw}{\textrm{pw}}
\newcommand{\vx}{\textrm{vx}}

\newcommand{\calA}{{\cal A}}

\newcommand{\YL}[1]{{\color{cyan} {#1}}}

\begin{document}

\title{Off-axis vortex scattering of electron-positron annihilation into a photon pair}

\author{Yi Liao}
\email{liaoy@m.scnu.edu.cn}
\affiliation{School of Physics, Nankai University, Tianjin 300071, China}
\affiliation{State Key Laboratory of Nuclear Physics and
Technology, Institute of Quantum Matter, South China Normal
University, Guangzhou 510006, China}
\affiliation{Guangdong Basic Research Center of Excellence for
Structure and Fundamental Interactions of Matter, Guangdong
Provincial Key Laboratory of Nuclear Science, Guangzhou
510006, China}
\author{Quan-Yu Wang}
\email{1120200055@mail.nankai.edu.cn}
\affiliation{School of Physics, Nankai University, Tianjin 300071, China}

\author{Yuanbin Wu}
\email{yuanbin@nankai.edu.cn}
\affiliation{School of Physics, Nankai University, Tianjin 300071, China}

\begin{abstract}

The off-axis triple-vortex scattering process of 
$e^-e^+\to\gamma\gamma$ is studied theoretically, in which the positron is in a plane-wave state and the electron and photons are in vortex states. We develop a theoretical formalism for the process, which allows us to study the effects of various vortex parameters and scattering angle. We adopt a Bessel-Gaussian type wave packet for the initial vortex electron for the purpose of normalization. Numerical calculations are performed for an electron and a positron with a moderate energy around $1~\textrm{MeV}$. Our results demonstrate strong impacts of the scattering angle and the topological charges on the cross section and distributions in the energy and cone angles of the vortex photons. This could provide insight into off-axis vortex scattering and also a possible approach to distinguishing and detecting vortex electrons by off-axis vortex scattering.

\end{abstract}

\maketitle

\section{Introduction}

The vortex state of particles, which carries an intrinsic orbital angular momentum with respect to its propagation direction, opens unprecedented opportunities for novel effects and applications \cite{Bliokh:2017uvr,Yuan:2017,Ivanov:2022jzh,Yao:11,shen:2019,Forbes:2024,Bliokh:2015doa,Knyazev:2018} in optics \cite{mair:2001,Wang:2015vrl,He:1995}, atomic physics and material science \cite{Picón:2010,Afanasev:2016,Serbo:2015kia,Karlovets:2016uhb,Karlovets:2015nva,Maiorova:2018inm,Mandal:2020ycl,Sarenac:2024tnj,madan:2020}, nuclear physics \cite{madan:2020,Larocque:2018,Wu:2021trm,Lu:2023wrf,Lu:2024gha}, and particle physics \cite{kaminer:2015,Groshev:2019uqa,Silenko:2019dfx,Hayrapetyan:2014faa,Bandyopadhyay:2015eri,Aleksandrov:2022fmp}. This benefits from the experimental demonstration and manipulation of vortex states \cite{Bliokh:2017uvr,Yuan:2017,Ivanov:2022jzh}. Starting from the introduction of optical vortex beams in 1992 \cite{Allen:1992zz}, vortex photons \cite{shen:2019,Forbes:2024,Heckenberg:1992,BEIJERSBERGEN:1994,Ruffato:2014,Longman:2020,Peele:2002,Terhalle:2011,Gariepy:2014,Hemsing:2013,Bahrdt:2013eoa,Gauthier:2016,Lee:2019} from visible light to x-ray, low-energy vortex electrons \cite{Ivanov:2022jzh,Uchida:2010hbm,Verbeeck:2010ezk,McMorran:2011bql,Mafakheri:2017,Tavabi:2022,Vanacore:2019} with a kinetic energy up to $300$ keV, cold vortex neutrons \cite{Clark:2015rcq,Sarenac:2019,Sarenac:2022}, and slow vortex atoms \cite{luski:2021} have been experimentally produced by various traditional techniques \cite{Bliokh:2017uvr,Yao:11,shen:2019,Yuan:2017,Ivanov:2022jzh,Forbes:2024} such as spiral phase plates and holographic gratings. Furthermore, the production of high-energy vortex photons and electrons has been proposed based on scattering processes, including Compton scattering \cite{Katoh:2016yqc,Petrillo:2016,Jentschura:2010ap,Jentschura:2011ih,Katoh:2016aww}, nonlinear Compton scattering \cite{Taira:2017,Chen:2018tkb,Ababekri:2022mob,Guo:2023uyu,Bogdanov:2019ocq,Jiang:2024fit,Bu:2024}, and laser-plasma interactions \cite{Wang:2020,liu:2016,hu:2021,ju:2018}, as well as the method of accelerating vortex electrons by electromagnetic fields \cite{Karlovets:2020tlg,Meng:2025isy,Murtazin:2025gic,Zmaga:2025haq,Karlovets:2025lnq} and the method via generalized measurements or quantum entanglement \cite{Karlovets:2022evc,Karlovets:2022mhb}, although the realization of high-energy vortex particles so far still remains a challenge \cite{Ivanov:2022jzh,Bu:2024,Li:2024mzd}.

The study of vortex scattering, which has been shown to be a novel method to probe quantum electrodynamics (QED) effects and nuclear and hadronic processes, is one of the new opportunities opened by the experimental demonstration of vortex particles \cite{Ivanov:2022jzh}. Furthermore, vortex-state QED scattering processes are believed to be possible ways to produce and detect high-energy vortex states \cite{Jentschura:2010ap,Jentschura:2011ih,Ababekri:2022mob,Guo:2023uyu,Bu:2024,Ababekri:2024cyd,Bu:2021ebc,Lei:2021eqe,Liao:2025skb}. However, so far the studies are mainly focused on the so-called on-axis case, i.e. when the outgoing vortex states move along the collision axis \cite{Ivanov:2022jzh,Jentschura:2010ap,Jentschura:2011ih,Liao:2025skb,Bu:2021ebc,Lei:2021eqe,Bu:2024,Ababekri:2022mob,Guo:2023uyu,Ababekri:2024cyd}. 
Even kinematic discussion of off-axis scattering is rather limited~\cite{Ivanov:2011tu}, and its dynamics remains to be explored.

In this work we set out to explore dynamics in off-axis vortex scattering by focusing on a specific QED process, i.e. the electron-positron annihilation into a photon pair. As one of the fundamental QED processes \cite{Laudau:1982,Sakurai:1968}, it has been shown to be important for astrophysical phenomena \cite{Aksenov:2008zz,Svensson:1982hz,Coppi:1990,Churazov:2005} as well as for testing entanglement of annihilation photons \cite{Ivashkin:2022qdr,Tkachev:2024iuj}. More concretely, we study the configuration of the triple-vortex process $e^-e^+\to\gamma\gamma$ in which the positron is in a plane-wave state and the electron and photons are in vortex states. We develop a theoretical formalism 
to calculate the process analytically. For the purpose of physical normalization, we adopt a Bessel-Gaussian (BG) type wave packet \cite{Ivanov:2011bv,Liao:2025skb,Sheremet:2024jky} 
to describe the initial vortex electron. We perform numerical 
analyses for an electron with a moderate energy of 1~MeV for which off-axis scattering is more relevant than at high energy, to assess the impact of various vortex parameters on off-axis vortex scattering.

This article is organized as follows. In \cref{sec:analytical}, we present the theoretical formalism for the analytical calculation of the off-axis triple-vortex scattering process. 
In \cref{sec:numerical}, we present our numerical results and investigate the influences of the scattering angle, the total angular momenta (TAMs) and cone angles of the vortex electron and photons. 
We finally conclude in \cref{sec:conclusion}. The natural units ($\hbar=c=1$) are used throughout, and $\alpha=e^2/(4\pi)\sim 1/137$ is the fine structure constant with $e$ being the electron charge.

\section{Analytical Calculation}
\label{sec:analytical}

In this section, we calculate the differential cross section for the triple-vortex process $e^-e^+\to\gamma\gamma$ in which the positron is in a plane-wave state and the electron and photons are in a vortex state. We start with the description of the simplest Bessel vortex and its normalization. To obtain a physically meaningful cross section in our later numerical analysis, we normalize the initial electron vortex state by a Gaussian form while keeping the final two photons in the simplest Bessel state. We then calculate the triple-vortex scattering amplitude in terms of the plane-wave amplitude, working out kinematics along the way. In the final subsection we give the formula for the differential cross section which will be employed for our numerical analysis in the next section.

\subsection{Vortex states and their normalization}

The wavefunction for an electron in a Bessel vortex state moving in the $z$ direction is, 
\begin{eqnarray}
\label{eq:Bessel_state}
\psi_{\kappa,k_{z};m,h_-}(x)
&=& e^{i(k_zz-E_kt)}\int\frac{d^2\kvec_\perp}{(2\pi)^2}a_{\kappa m}(\kvec_\perp)u(k,h_-)
e^{i\kvec_\perp\cdot\xvec_\perp},
\end{eqnarray}
where $k_z,~\kappa$ are the $z$-component of the momentum $\kvec$ and magnitude of the transverse momentum $\kvec_\perp$ respectively, and $E_k$ the electron energy. The electron has a half-integer topological charge $m$ which is the projection of its TAM in the propagation direction of the vortex. The plane-wave bispinor $u(k,h_-)$ with a definite helicity $h_-/2$ is given in \cref{eq:bispinor}, and the Bessel kernel is,  
\begin{eqnarray}
\label{eq:BesselKernel}
a_{\kappa m}(\kvec_\perp) 
&=& i^{-m} e^{im\varphi} \sqrt{\frac{2\pi}{\kappa}} \delta(k_\perp-\kappa),
\end{eqnarray}
where $\varphi$ is the azimuthal angle of $\kvec_\perp$ and $k_\perp=|\kvec_\perp|$. A vortex photon of energy $\omega$ moving in the $z$ direction has a similar wavefunction, 
\begin{eqnarray}
\vecA_{\kappa,k_z;m,\lambda}(x) 
&=& e^{i(k_zz-\omega t)} \int\frac{d^2\kvec_\perp}{(2\pi)^2}a_{\kappa m}(\kvec_\perp) \vec\epsilon(k,\lambda) e^{i\kvec_\perp\cdot\xvec_\perp},
\end{eqnarray}
where $m$ is an integer topological charge, and the polarization vector $\vec\epsilon(k,\lambda)$ for helicity $\lambda$ is given in \cref{eq:polarization}. 
As we will consider off-axis scattering in \cref{sec:amplitude} where the vortex photons in the final state do not share the same vortex axis as the initial vortex electron, it is important to notice that the topological charge of each vortex is measured with respect to its own propagation direction, thus also dubbed orbital helicity~\cite{Ivanov:2011tu}. 

Now we consider the normalization and related issues of a Bessel vortex state. According to discussions in \cite{Ivanov:2011kk,Jentschura:2011ih} as detailed in \cref{sec:appendix_A}, the normalization factors for the initial vortex electron and final vortex photons are:
\begin{subequations}
\begin{eqnarray}
N_{e,\vx}&=&\frac{1}{\sqrt{2E_k}}\sqrt{\frac{\pi}{RL_{z}}},
\\
N_{\gamma_{a},\vx}&=&\frac{1}{\sqrt{2\omega_{a}}}\sqrt{\frac{\pi}{RL_{z'}}},
\end{eqnarray}
\end{subequations}
where $L_z,~L_{z'}$ are the large lengths in the $z$ and $z'$ propagation directions, $R$ is a common large radius in the transverse planes, and $a=1,~2$ enumerates the two photons in the final state. The phase-space measure for the two vortex photons is 
\begin{equation}
d\textrm{PS}=\left(\frac{RL_{z'}}{\pi}\right)^2 \frac{dk_{1\perp'}dk_{1z'}}{2\pi}\frac{dk_{2\perp'}dk_{2z'}}{2\pi},
\end{equation}
where a prime indicates transverse and longitudinal components in the final-state $x'y'z'$ frame; see \cref{sec:amplitude}, \cref{sec:appendix_B}, and \cref{fig:HLW}.

Since the Bessel vortex of the initial electron causes the non-integrable problem, we will replace it by a Bessel-Gaussian vortex while leaving the final photons in a Bessel vortex for simplicity. The Bessel-Gaussian vortex is defined as a smearing integration of $\psi_{\kappa,k_{z};m,h_-}(x)$ in Eq.~\eqref{eq:Bessel_state} weighted by a Gaussian function:
\begin{eqnarray}
\psi_{\kappa_0\sigma_\perp;k_{z}\sigma_{z};m,h_-}^{\BG}(x)
&=&\frac{1}{2\pi}\int^{\infty}_{-\infty} dk_\parallel\int_0^\infty d\kappa~\psi_{\kappa,k_\parallel;m,h_-}(x) f(\kappa,k_\parallel;\kappa_0,k_{z},\sigma_\perp,\sigma_{z}),
\label{eq:initial_eBG}
\end{eqnarray}
where the normalized Gaussian function is
\begin{eqnarray}
f(\kappa,k_\parallel;\kappa_0,k_{z},\sigma_\perp,\sigma_{z})
=N(\kappa_0,\sigma_\perp)N(\sigma_z)
\exp\bigg[-\frac{(\kappa-\kappa_0)^2}{2\sigma_\perp^2}\bigg]\exp\bigg[-\frac{(k_\parallel-k_z)^2}{2\sigma_z^2}\bigg],
\label{eq:gaussian}
\end{eqnarray}
and the normalization factors are determined by 
\begin{subequations}
\begin{eqnarray}
N^{-2}(\kappa_0,\sigma_\perp)&=&
\frac{1}{2}\sqrt{\pi}\sigma_\perp \left[1+\textrm{erf}\left(\frac{\kappa_0}{\sigma_\perp}\right)\right],
\\
N^{-2}(\sigma_z)&=&\frac{\sigma_z}{2\sqrt{\pi}}.
\end{eqnarray}
\end{subequations}
The independent smearing of longitudinal and transverse momentum implies that the vortex state is not exactly mono-chromatic. Although this is experimentally more practical, we do this mainly for theoretically removing normalization factors associated with the spatial sizes $R,~L_z$. By imposing the relation between the transverse and longitudinal smearing widths, 
\begin{equation}
\label{width_eq}
k_z\sigma_z=\kappa\sigma_\perp,
\end{equation}
the electron vortex is almost mono-chromatic since $\sigma_z=\sigma_\perp\tan\theta$ is not large for a relatively small cone angle $\theta$. 
Normalizing the Bessel-Gaussian wavefunction to one particle in the whole space and applying the orthogonality condition of the Bessel function 
\begin{equation}
\int_{0}^{\infty}xdx~J_{m}(\kappa{x})J_{m}(\tilde\kappa{x})=\frac{1}{\kappa}\delta(\kappa-\tilde\kappa),
\end{equation}
the normalization factor of the Bessel-Gaussian vortex is determined to be 
\begin{equation}
\label{eq:Norm_NBG}
N_{\textrm{BG}}=\frac{1}{\sqrt{2E_{k}}}.
\end{equation}

\subsection{Scattering amplitude}
\label{sec:amplitude}

\begin{figure}
  \centering
  \includegraphics[width=10.0cm]{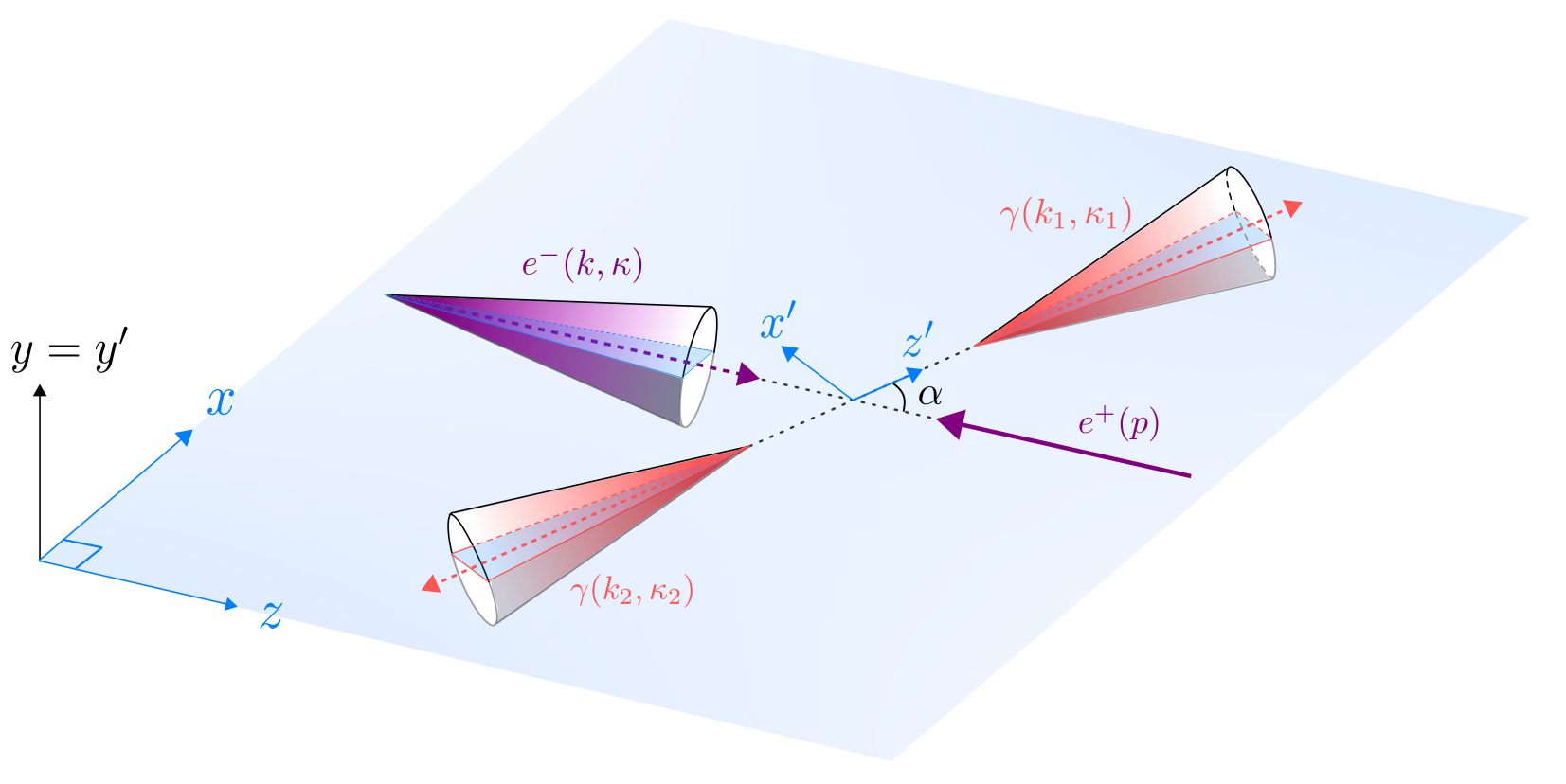}
  \\
  \caption{Scattering geometry.}
  \label{fig:HLW}
\end{figure}

We are going to calculate the amplitude for the triple-vortex scattering process: 
\begin{eqnarray}
e^-(k,\kappa,m,h_-) +e^+(p,h_+)\to
\gamma(k_1,\kappa_1,m_1,\lambda_1) +\gamma(k_2,\kappa_2,m_2,\lambda_2),
\label{eq:triple_vortex}
\end{eqnarray}
where $k,~p,~k_a$ ($a=1,~2$) are the magnitudes of the three-momenta $\kvec,~\pvec,~\kvec_a$ and $\kappa,~\kappa_a$ are the magnitudes of transverse momenta of the vortex states, so that the sines of the cone angles are $\sin\theta=\kappa/k$ and $\sin\theta_a=\kappa_a/k_a$ respectively. Following \cite{Ivanov:2011tu}, we call $m,~m_a$ the orbital helicities of the electron and photons, i.e. the projections of their TAMs in their respective propagation directions although the adjective orbital is a bit confusing for a particle with spin. And $h_\mp/2$ and $\lambda_a$ are the helicities; for a vortex state, this means that each of its plane-wave components is in the same helicity state although there is no way to define a helicity for the overall vortex state itself. We consider the scattering geometry as shown in \cref{fig:HLW} where the vortex electron propagating in the $z$ axis annihilates a plane-wave positron moving in the $-z$ axis to produce a pair of vortex photons propagating back-to-back making an angle $\alpha$ with respect to the $z$ axis. To make life easier, we can always choose a common $y=y'$ axis for the initial-state $xyz$ and final-state $x'y'z'$ frames. For simplicity, we work in the reference frame where $k_z+p_z=0$ with $k_z=k\cos\theta>0$ and $p_z=-p<0$. Note that we cannot generally define a center-of-mass frame in the usual sense when a vortex state is involved. The polar and azimuthal angles $(\theta,\varphi)$ of the vortex electron are defined in the $xyz$ frame, while those of the vortex photons $(\theta_a,\varphi_a)$ are in the $x'y'z'$ frame.

The scattering matrix element for the triple-vortex process in \eqref{eq:triple_vortex} is 
\begin{eqnarray}
&&S_\vx=\iiint\frac{d^2\kvec^\perp}{(2\pi)^2}\frac{d^2\kvec_1^\perp}{(2\pi)^2}\frac{d^2\kvec_2^\perp}{(2\pi)^2}
a_{\kappa m}(\kvec^\perp)a^*_{\kappa_1 m_1}(\kvec_1^\perp)
a^*_{\kappa_2 m_2}(\kvec_2^\perp)S_\pw,
\label{eq:S_matrix_def}
\end{eqnarray}
where $\kvec^\perp,~\kvec_a^\perp$ are the transverse momenta of the respective vortices, $a_{\kappa m}(\kvec^\perp)$ for a Bessel vortex is defined in Eq.~\eqref{eq:BesselKernel}, and the scattering matrix element for the same process but with all particles being in a plane-wave state is 
\begin{eqnarray}
&&S_\pw=i(2\pi)^4\delta^4(k+p-k_1-k_2)\calA_\pw,
\end{eqnarray}
where $\calA_\pw$ is the usual plane-wave invariant amplitude which is given in \cref{sec:appendix_B} for arbitrary momenta and helicities. Note that we also use $k$ etc to denote a four-momentum. While the magnitudes of transverse momenta $k^\perp=|\kvec^\perp|,~k_a^\perp=|\kvec_a^\perp|$ are fixed to be $\kappa,~\kappa_a$ by the $\delta$ functions in the Bessel kernels, the azimuthal angles $\varphi,~\varphi_a$ can be finished using the $\delta$ functions for three-momentum conservation. Denoting 
\begin{eqnarray}
q=k_{1z'}+k_{2z'}
=\omega_1\cos\theta_1-\omega_2\cos\theta_2,
\end{eqnarray}
conservation of the $z'$-component momenta gives 
\begin{eqnarray}\label{q_delta}
\delta(\kappa\sin\alpha\cos\varphi-q)
=\frac{1}{\kappa\sin\alpha}\delta\Big(\cos\varphi-\frac{\sin\chi}{\sin\alpha}\Big),
\label{eq:z-component}
\end{eqnarray}
where we define 
\begin{eqnarray}
\sin\chi=\frac{q}{\kappa},
\end{eqnarray}
and choose $|\chi|\leq\pi/2$ without loss of generality so that $\cos\chi\geq 0$. Denoting the planar vector 
\begin{eqnarray}
\Kvec=\kvec_1^\perp+\kvec_2^\perp,
\label{eq:triangle}
\end{eqnarray}
with the azimuthal angle $\Phi$ in the $x'y'z'$ frame and magnitude $K$, conservation of the planar components of momenta, i.e. $\kappa\cos\alpha\cos\varphi-K\cos\Phi=0$ and $\kappa\sin\varphi-K\sin\Phi=0$, then yields, 
\begin{eqnarray}
K=\kappa\cos\chi,\quad
\cos\Phi=\frac{\tan\chi}{\tan\alpha},
\label{eq:xy-component}
\end{eqnarray}
where \eqref{eq:z-component} has been applied. We will indicate by a star the values of $\varphi,~\Phi$ fulfilling momentum conservation in \eqref{eq:z-component} and \eqref{eq:xy-component}, i.e.
\begin{eqnarray}
\cos\varphi^*=\frac{\sin\chi}{\sin\alpha},\quad
\cos\Phi^*=\frac{\tan\chi}{\tan\alpha}.
\end{eqnarray}
Note that there are two solutions to $\varphi^*\in[0,2\pi)$. Choosing arbitrarily one of them $\varphi^*\in[0,\pi)$ so that $\sin\varphi^*\geq 0$ by default below, the other solution is $(2\pi-\varphi^*)\in[\pi,2\pi)$. Correspondingly, there are two solutions to $\Phi^*$, which we also denote by $\Phi^*$ and $2\pi-\Phi^*$. The original four-momentum conservation $\delta$ functions become 
\begin{eqnarray}
\delta^4(k+p-k_1-k_2)
=\frac{1}{\kappa\sin\alpha}\delta(E-\omega_1-\omega_2)
\delta(\cos\varphi - \cos\varphi^*)
\delta^2(\kvec_1^\perp+\kvec_2^\perp-\Kvec),
\end{eqnarray}
where $E=E_k+E_p$, and $E_{k,p}$, $\omega_a=k_a$ are the energies of the electron, positron, and photons. 

To integrate over azimuthal angles $\varphi,~\varphi_a$, we explicitly display their dependence in the plane-wave amplitude $\calA_\pw(\varphi,\varphi_1,\varphi_2)$. Collecting the phases from the Bessel kernels, the $\varphi$ integral is trivially finished using the $\delta$ function: 
\begin{eqnarray}
I&=&
\iiint d\varphi d\varphi_1 d\varphi_2~
\delta(\cos\varphi - \cos\varphi^*)
\delta^2(\kvec_1^\perp+\kvec_2^\perp-\Kvec)
F(\varphi,\varphi_1,\varphi_2)
\nonumber
\\
&=&
\frac{1}{\sin\varphi^*}\iint d\varphi_1 d\varphi_2~
\left.\delta^2(\kvec_1^\perp+\kvec_2^\perp-\Kvec)\right|_{\varphi=\varphi^*}
\Big[F(\varphi^*,\varphi_1,\varphi_2)
+F(2\pi-\varphi^*,\varphi_1,\varphi_2)\Big],
\end{eqnarray}
where 
\begin{eqnarray}
&&F(\varphi,\varphi_1,\varphi_2)
=\calA_\pw(\varphi,\varphi_1,\varphi_2)
e^{i[m\varphi-(m_1\varphi_1\YL{-}m_2\varphi_2)]}.
\end{eqnarray}
The planar integrals have been worked out in \cite{Ivanov:2011kk,Liao:2025skb}: 
\begin{eqnarray}
\iint d\varphi_1d\varphi_2~
\left.\delta^2(\kvec_1^\perp+\kvec_2^\perp-\Kvec)\right|_{\varphi=\varphi^*}
F(\varphi^*,\varphi_1,\varphi_2)
&=&
\frac{1}{2\Delta}\sum_{\eta=\pm}F(\varphi^*,\Phi^*-\eta\delta_1,\Phi^*+\eta\delta_2),
\end{eqnarray}
where $\delta_{1,2}$ are the inner angles of the triangle formed by $\kvec_a^\perp$ and $\Kvec$, and $\Delta$ is its area:  
\begin{eqnarray}
&&\delta_{1,2}
=\arccos\frac{K^2+\kappa_{1,2}^2-\kappa_{2,1}^2}{2K\kappa_{1,2}},\quad
\Delta
=\frac{1}{4}\sqrt{-\lambda(\kappa_1^2,\kappa_2^2,K^2)},
\end{eqnarray}
where $\lambda(a,b,c)=a^2+b^2+c^2-2ab-2bc-2ca$. 
In summary, the integration over azithumal angles yields 
\begin{eqnarray}
\label{eq:general_amp}
I&=&\frac{1}{2\Delta\sin\varphi^*}\sum_{\eta=\pm}
e^{i\eta(m_1\delta_1+m_2\delta_2)}
\Big[e^{i(m\varphi^*-m_1\Phi^*+m_2\Phi^*)}
\calA_\pw\left(\varphi^*,\Phi^*-\eta\delta_1,\Phi^*+\eta\delta_2\right)
\nonumber
\\
&&+e^{i2\pi(m-m_1+m_2)}e^{-i(m\varphi^*-m_1\Phi^*+m_2\Phi^*)}
\calA_\pw\big(2\pi-\varphi^*,2\pi-\Phi^*-\eta\delta_1,2\pi-\Phi^*+\eta\delta_2\big)\Big].
\end{eqnarray}
The scattering matrix in \eqref{eq:S_matrix_def} becomes 
\begin{subequations}
\begin{eqnarray}
\label{eq:general_S}
S_\vx&=&
i\delta(E-\omega_1-\omega_2)\calA_\vx,
\\
\calA_\vx&=&
i^{m_1+m_2-m}\frac{1}{\sin\alpha}
\sqrt{\frac{\kappa_1\kappa_2}{2\pi\kappa}}I.
\end{eqnarray}
\end{subequations}
When one considers $\calA_\pw$ as an azimuthal constant $\calA_0$ and ignores spins, the phase $e^{i2\pi(m-m_1+m_2)}=1$ since $m,~m_a$ are integers, so that the azimuthal integral simplifies to 
\begin{eqnarray}
I&\to&\frac{2}{\Delta\sin\varphi^*}
\cos(m_1\delta_1+m_2\delta_2)
\cos(m\varphi^*-m_1\Phi^*+m_2\Phi^*)\calA_0.
\end{eqnarray}
Using $\sin\varphi^*\sin\alpha=\sqrt{\sin^2\alpha-\sin^2\chi}$, the kinematical result in \cite{Ivanov:2011tu} is reproduced: 
\begin{subequations}
\begin{eqnarray}
S_\vx&\to&i\delta(E-\omega_1-\omega_2)
i^{m_1+m_2-m}\calA_\textrm{B},
\\
\label{Ivan_S}
\calA_\textrm{B}&=&\frac{2}{\Delta}
\sqrt{\frac{\kappa_1\kappa_2}{2\pi\kappa}}
\frac{\cos(m_1\delta_1+m_2\delta_2)
\cos(m\varphi^*-m_1\Phi^*+m_2\Phi^*)}{\sqrt{\sin^2\alpha-\sin^2\chi}}\calA_0.
\end{eqnarray}
\end{subequations}

\subsection{Differential cross section}

Since the ideal Bessel vortex is not square-integrable, it results in a non-integrable singularity in cross section through the factor $\Delta^{-2}$ which is accompanied by a normalization factor proportional to $(RL_z)^{-1}$. The simplest approach to address this issue is to assume a Bessel-Gaussian vortex for the initial electron while still adopting a Bessel vortex for the final-state photons. Furthermore, since the initial positron is in a plane-wave state that is uniform in space, the flux is not affected by the Gaussian smearing, i.e. still proportional to $|v_--v_+|$, where $v_-=k_z/E_k=(k/E_k)\cos\theta$ and $v_+=p_z/E_p=-(k/E_p)\cos\theta$ are the propagation velocities of the electron and positron in the $z$ axis. The differential cross section is 
\begin{equation}
\label{BG_sigma}
d^4\sigma=\frac{1}{2E_{p}}\frac{N_{\textrm{BG}}^2}{|v_- -v_+|} |\calA^{\textrm{BG}}_\vx|^2\frac{1}{2\pi}\delta(E-\omega_1-\omega_2)
\frac{d\kappa_{1}dk_{1z}}{2\omega_{1}2\pi}
\frac{d\kappa_{2}dk_{2z}}{2\omega_{2}2\pi},
\end{equation}
where the primes on variables of photons in the $x'y'z'$ frame have been suppressed for brevity and 
\begin{equation}
\label{eq:BG_amplitude}
\calA^{\textrm{BG}}_\vx=\frac{1}{{2\pi}}\int^{\infty}_{-\infty}dk_\parallel\int_0^\infty d\kappa~f(\kappa,k_\parallel;\kappa_0,k_{z},\sigma_\perp,\sigma_{z})\calA_\vx.
\end{equation}
The differential cross section can also be expressed in terms of the energies and cone angles of the vortex photons as 
\begin{eqnarray}
\label{eq:cross_section_polar_angle}
d^4\sigma = \frac{1}{16E_{k}E_{p}}\frac{1}{(2\pi)^3} \frac{1}{|v_- -v_+|} 
|\calA^{\textrm{BG}}_\vx|^2 \delta(E-\omega_1-\omega_2) d\omega_1d\omega_2d\theta_1d\theta_2,
\end{eqnarray}
where one of the energies, say $\omega_2$, can be finished using the $\delta$ function:
\begin{subequations}
\begin{eqnarray}
\label{eq:3d_sigma}
\frac{d^3\sigma}{d\omega_1d\theta_1d\theta_2}
&=&G(\omega_1;\theta_1,\theta_2),
\\
\label{eq:Gfunc}
G(\omega_1;\theta_1,\theta_2)
&=&\frac{1}{16E_{k}E_{p}}\frac{1}{(2\pi)^3} 
\frac{1}{|v_- -v_+|}|\calA^{\textrm{BG}}_\vx|^2,
\end{eqnarray}
\end{subequations}
where $\omega_2=E-\omega_1$ is implied in the function $G$.

The allowed ranges of the kinematical variables of the photons are worked out by energy-momentum conservation. Expressing $k\cdot{k_1}=p\cdot{k_2}$ in terms of the scalar products of three-momenta and working in our reference frame with $k_z+p_z=0$, energy conservation yields 
\begin{eqnarray}
\label{eq:energy_photon}
\frac{\omega_1}{E_k+E_p}=\frac{E_p-kR_2 c_\theta}
{(E_k+E_p)-k(R_1+R_2 c_\theta)},
\end{eqnarray}
where $R_1=\hat{\kvec}\cdot\hat{\kvec}_1$ and $R_2=\hat{\pvec}\cdot\hat{\kvec}_2$ are expressed in terms of various angles: 
\begin{subequations}
\begin{eqnarray}
R_1&=&s_{\theta_1}[s_\theta
(c_\alpha c_\varphi c_{\varphi_1} +s_\varphi s_{\varphi_1}) 
-c_\theta s_\alpha c_{\varphi_1}] 
+c_{\theta_1}[s_\alpha s_\theta c_\varphi
+c_\alpha c_\theta],
\\
R_2&=& 
s_\alpha s_{\theta_2}c_{\varphi_2} -c_\alpha c_{\theta_2},
\end{eqnarray}
\end{subequations}
with the shortcuts $c_\beta=\cos\beta$ and $s_\beta=\sin\beta$ for $\beta=\alpha,~\theta,~\varphi,~\theta_a,~\varphi_a$. 
Note that the azimuthal angles $\varphi,~\varphi_{1,2}$ have been fixed in \cref{sec:amplitude} in terms of other variables such as the magnitudes of transverse momenta and the auxiliary angle $\chi$. As also detailed in \cref{sec:amplitude}, momentum conservation in the $z'$ axis implies a restriction $\sin\alpha>|\sin\chi|$, while momentum conservation in the $x'y'$ plane demands $\kvec_a^\perp$ and $\Kvec$ to form a triangle \eqref{eq:triangle}, i.e. the sum of any two out of $\kappa_a$ and $|\kvec^\perp|\cos\chi$ is larger than the other. These restrictions together set the allowed ranges of the energies and cone angles of the photons.

\section{Numerical Analysis}\label{sec:numerical}

In our numerical analysis we assume a moderate energy with the central value $1~\textrm{MeV}$ for the vortex electron. We work in the reference frame where the vortex electron collides head-on against the plane-wave positron with $p_z=-k_z$ as described in \cref{sec:amplitude}. The transverse-momentum spread $\sigma_\perp$ of the BG electron is assumed to be half its central value $\kappa_0$, i.e., $\sigma_{\perp}=\kappa_{0}/2$, while the longitudinal-momentum spread $\sigma_z$ is then determined by \eqref{width_eq}. We assume two values of the electron TAM projection, $m=3/2,~21/2$, to study its impact, while 
two scattering angles $\alpha=\pi/6,~2\pi/3$ are selected for comparison.

\begin{figure}[H]
    \centering
    \includegraphics[width=0.5\linewidth]{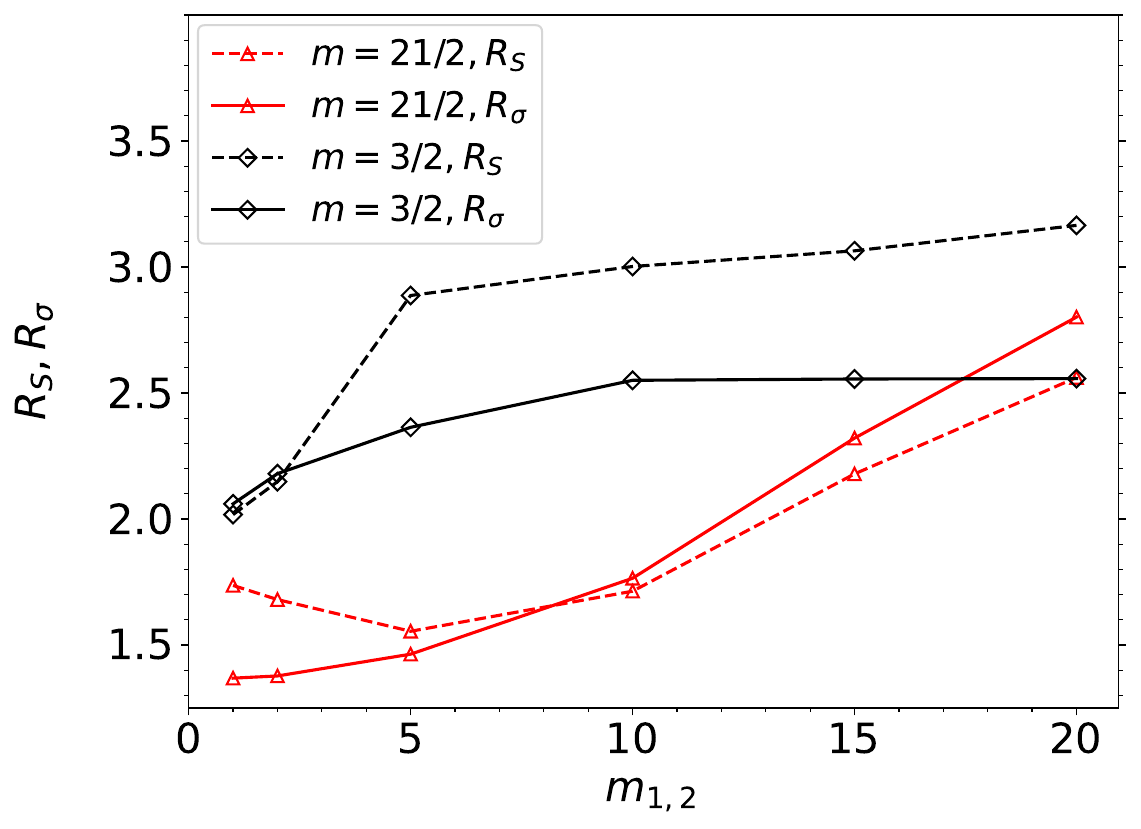}
    \caption{Comparison of $R_\sigma,~R_S$ variation in $m_1=m_2$ at fixed $\theta=\pi/6$ and $m=3/2,~21/2$.}
    \label{sigma_ratio}
\end{figure}

Before we embark on presenting our numerical results, we compare our result containing QED dynamics with the one in \cite{Ivanov:2011tu} involving only kinematical factors for a constant scattering amplitude of scalar particles. To simplify the matter in the following, we will consider some slices of multiple-variable space of the final photons; in particular, we will focus on two special cases of the cone angles of the final photons, i.e. we assume either $\theta_1=\theta_2=\theta_\gamma$ or when one of $\theta_{1,2}$ is fixed we study variations in the other. We define 
\begin{eqnarray}
\label{eq:sigma_gamma}
\sigma_\gamma&=&\int d\omega_1\int d\theta_\gamma~
G(\omega_1;\theta_\gamma,\theta_\gamma), 
\end{eqnarray}
where the function $G$ is defined in \eqref{eq:Gfunc}. Note that $\sigma_\gamma$ is not the total cross section but has the units of $\textrm{b/rad}$. For the purpose of comparison we remove common kinematical factors by taking the ratio of $\sigma_\gamma$ evaluated at two values of the scattering angle $\alpha$: 
\begin{eqnarray}
R_\sigma &=&
\frac{\sigma_\gamma(\alpha=\pi/6)}{\sigma_\gamma(\alpha=2\pi/3)}.
\end{eqnarray}
Similarly, by smearing the result in \eqref{Ivan_S} with a Gaussian function, 
\begin{eqnarray}
S&\propto&
\int dk_\parallel\int d\kappa~
f(\kappa,k_\parallel;\kappa_0,k_z,\sigma_\perp,\sigma_z) 
\calA_\textrm{B},
\end{eqnarray}
we define $\sigma_S$ in a manner similar to $\sigma_\gamma$ as 
\begin{eqnarray}
\sigma_S&\propto& \int d\omega_1\int d\theta_\gamma~
|S(\omega_1;\theta_\gamma,\theta_\gamma)|^2,
\end{eqnarray}
where $S$ is evaluated at $\theta_1=\theta_2=\theta_\gamma$ which is integrated in $\sigma_S$. And the ratio without involving QED dynamics is 
\begin{eqnarray}
R_S&=&\frac{\sigma_S(\alpha=\pi/6)}{\sigma_S(\alpha=2\pi/3)}.
\end{eqnarray}
In \cref{sigma_ratio} the ratios $R_\sigma$ and $R_S$ are shown as a function of the common TAM $m_1=m_2$ of the two photons at the fixed values of the electron vortex parameters ($\theta=\pi/6$ and $m=3/2,~21/2$). Our cross section has been summed (averaged) over the final (initial) state helicities $\lambda_{1,2}$ ($h_\pm/2$). 
We observe that whether the two ratios approach each other depends on the difference between $m$ and $m_1+m_2$ although there is no conservation law for orbital helicities in off-axis scattering. For $m_1=m_2$, while the $m$ and $m_{1,2}$ dependencies factorize in \eqref{Ivan_S} for pure kinematics, they are modulated by the plane-wave amplitude with built-in QED dynamics in \eqref{eq:general_amp}. This yields the general mismatched behavior of the two ratios in \cref{sigma_ratio}.

\begin{figure}[H]
    \centering
    \includegraphics[width=0.4\linewidth]{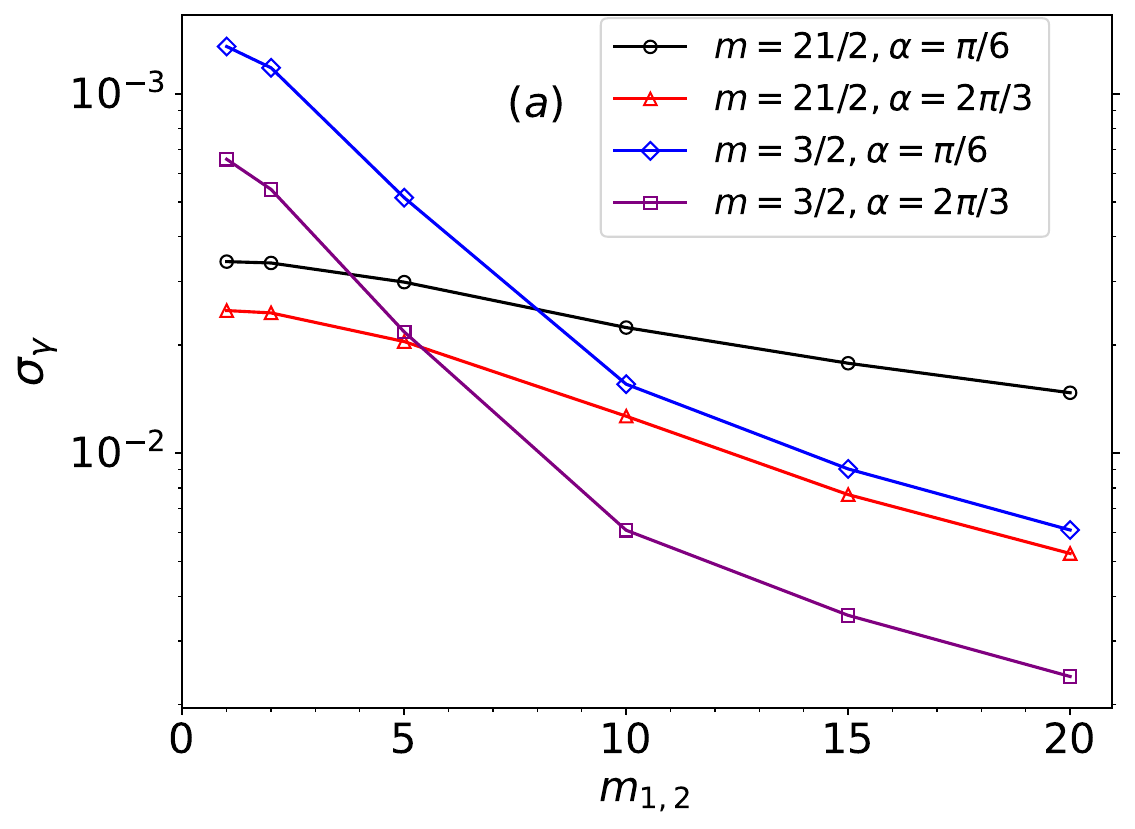}
    \includegraphics[width=0.4\linewidth]{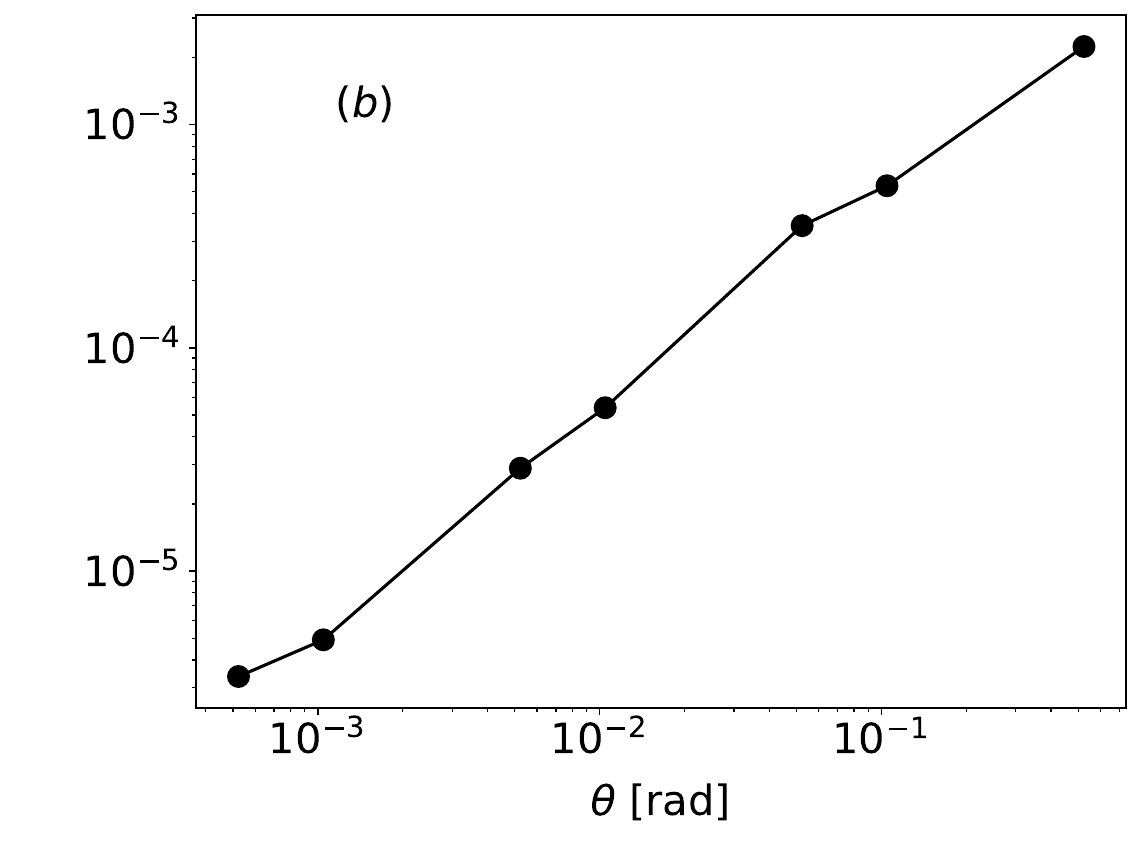}
\caption{$\sigma_{\gamma}\textrm{ (b/rad)}$ is shown as a function of $m_1=m_2$ in panel (a) with fixed $\theta=\pi/6$, $m=3/2,~21/2$, and $\alpha=\pi/6,~2\pi/3$, and as a function of $\theta$ in panel (b) with fixed $m=21/2$, $m_1=m_2=10$, and $\alpha=\pi/6$.}
\label{TAM_&_angle}
\end{figure}

In \cref{TAM_&_angle} we show $\sigma_\gamma$ in \eqref{eq:sigma_gamma} as a function of $m_1=m_2$ in panel (a) and a function of $\theta$ in panel (b). We observe that $\sigma_\gamma$ decreases sharply with the increase of $m_{1,2}$ for the case of $m=3/2$, while it drops more smoothly for the case of $m=21/2$. This clearly shows the impact of the initial electron TAM on the process. Moreover, $\sigma_\gamma$ never vanishes even for a large $m_{1,2}$. This reflects the fact that there is no conservation of orbital helicities or TAMs in off-axis scattering. This is in contrast to the case of on-axis scattering where projection of TAMs in the common axis is conserved. \cref{TAM_&_angle}(b) shows that $\sigma_\gamma$ is an increasing function of the open angle $\theta$ of the vortex electron. This behavior may be understood as follows. For a given scattering angle $\alpha$, unlike the case of plane-wave scattering in which the two photons have a definite energy, the energy of the vortex photon under consideration spans a region according to \eqref{eq:energy_photon}, which in turn depends strongly on $\theta$ of the vortex electron and results in a larger $\sigma_\gamma$.

\begin{figure}[H]
    \centering
    \includegraphics[width=0.6\linewidth]{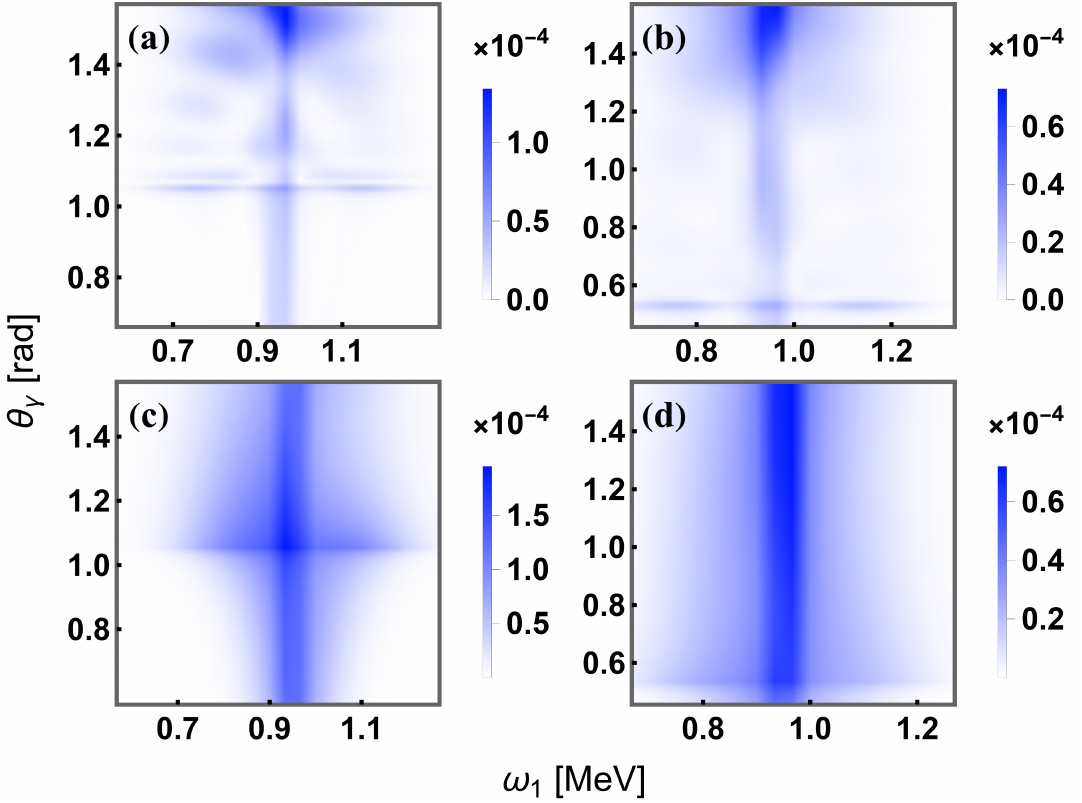}
    \caption{Differential cross section $G$ ($\textrm{b}/(\textrm{MeV }\textrm{rad}^2)$) in the plane $\theta_1=\theta_2=\theta_\gamma$ is shown as a density distribution in $\omega_1$ and $\theta_\gamma$. In all panels $m_{1}=m_{2}=1$ and $\theta=\pi/6$, and in panel (a,b,c,d), $(m,\alpha)=(21/2,\pi/6),~(21/2,2\pi/3),~(3/2,\pi/6),~(3/2,2\pi/3)$ respectively.}
    \label{2d_m1}
\end{figure}

\begin{figure}[H]
    \centering
    \includegraphics[width=0.6\linewidth]{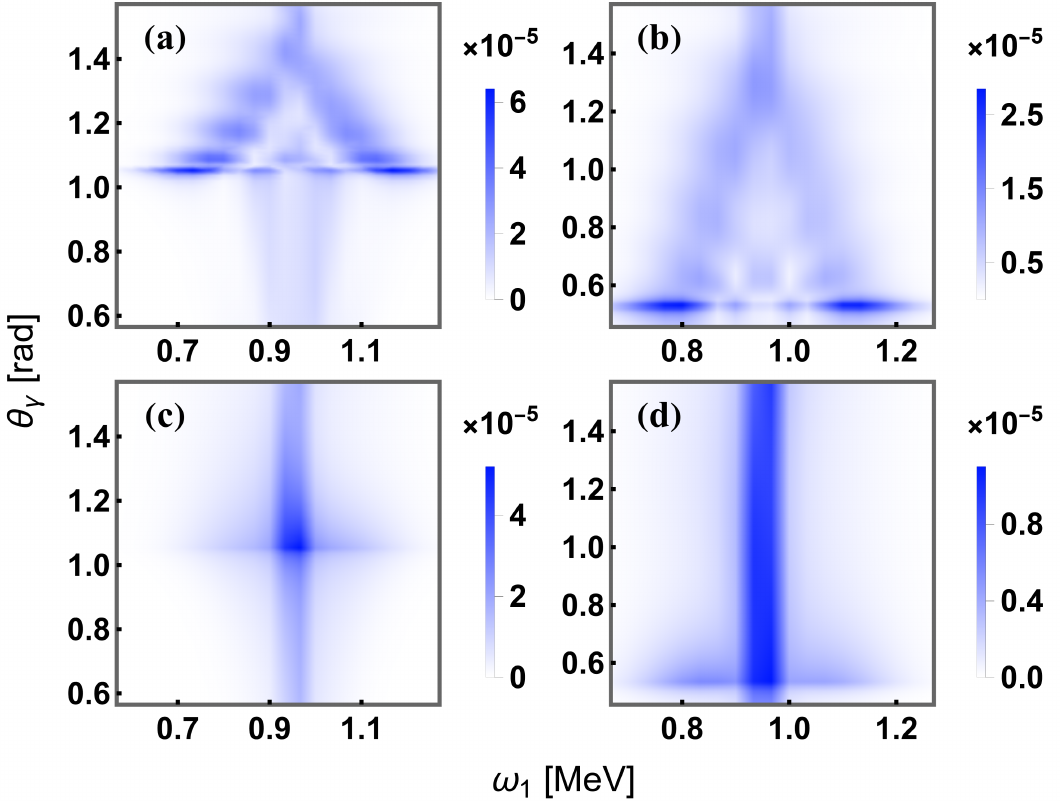}
    \caption{Same as \cref{2d_m1} except that $m_1=m_2=10$.}
    \label{2d_m10}
\end{figure}

We now turn to analyse the dependence of cross section on the cone angles and energies of the final vortex photons. For simplicity, we illustrate the three-dimensional differential cross section $G(\omega_1;\theta_1,\theta_2)$ in a special plane, $\theta_1=\theta_2=\theta_\gamma$, i.e. where the two photons share the same cone angle. We present in Fig.~\ref{2d_m1} and Fig.~\ref{2d_m10} the density distribution $G(\omega_1;\theta_\gamma,\theta_\gamma)$ in $\omega_1$ and $\theta_\gamma$ for selected choices of parameters. It can be seen that the differential cross section depends strongly on the topological charges of the photons $m_{1,2}$ (by comparing Fig.~\ref{2d_m1} and Fig.~\ref{2d_m10}) and of the electron $m$ and the scattering angle $\alpha$ (by comparing panels within one figure). The pattern of the differential cross section depends more on the topological charge $m$ and the scattering angle $\alpha$, while the characteristics of a pattern become more discernable with the increase of $m_{1,2}$. This implies that one may distinguish the topological charge $m$ of the initial vortex electron from the pattern of differential cross section at a given scattering angle.

\begin{figure}[H]
    \centering
    \includegraphics[width=0.6\linewidth]{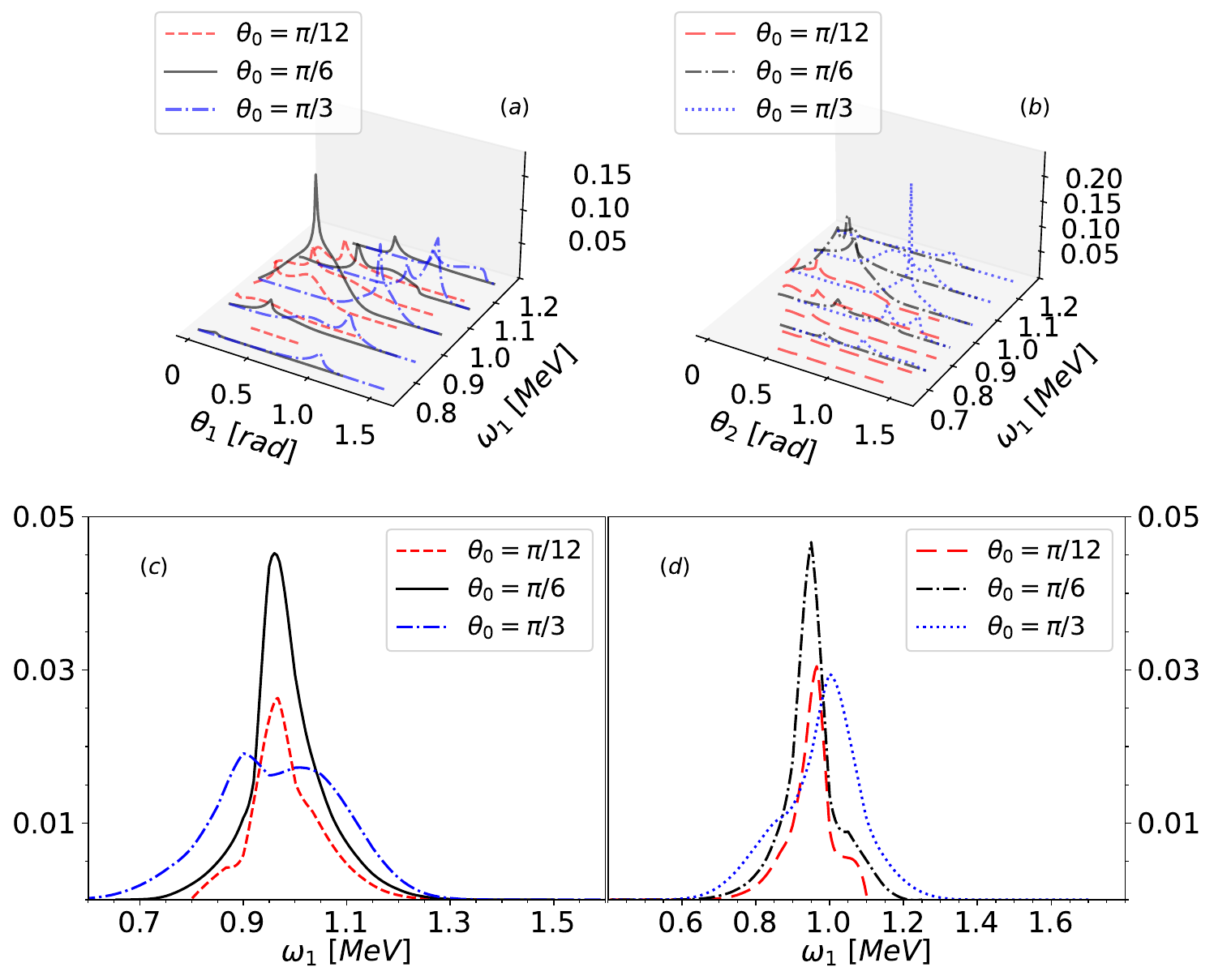}
    \caption{(a) Distribution $G_1(\omega_1;\theta_1)$ ($\textrm{b}/(\textrm{MeV }\textrm{rad}^2)$) in $\theta_1$ is shown at various values of $\omega_1$ and $\theta_2=\theta_0$. 
    (b) Similar to (a) but for $G_2(\omega_1;\theta_2)$. 
    (c) Distribution $H_1(\omega_1)$ (b/(MeV rad)) in $\omega_1$. 
    (d) Similar to (c) but for $H_2(\omega_1)$. 
    In all panels $m=3/2$, $m_1=m_2=1$, $\theta=\pi/6$, and $\alpha=\pi/6$.}
    \label{fig_angle}
\end{figure}

\begin{figure}[H]
    \centering
    \includegraphics[width=0.6\linewidth]{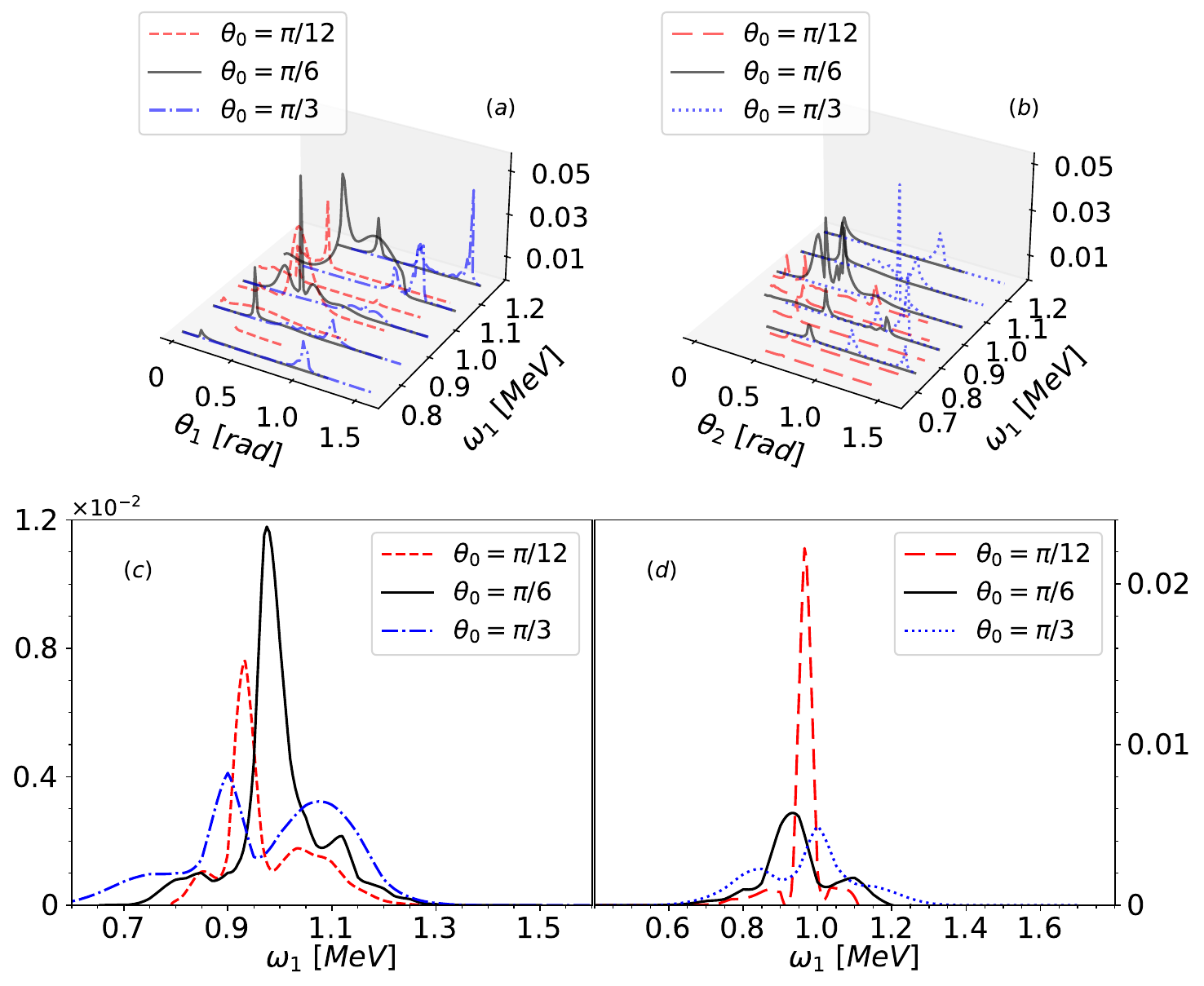}
    \caption{Same as Fig.~\ref{fig_angle} except that $m=21/2$ and $m_1=m_2=10$.}
        \label{fig_angle10}
\end{figure}

\begin{figure}[H]
    \centering
    \includegraphics[width=0.6\linewidth]{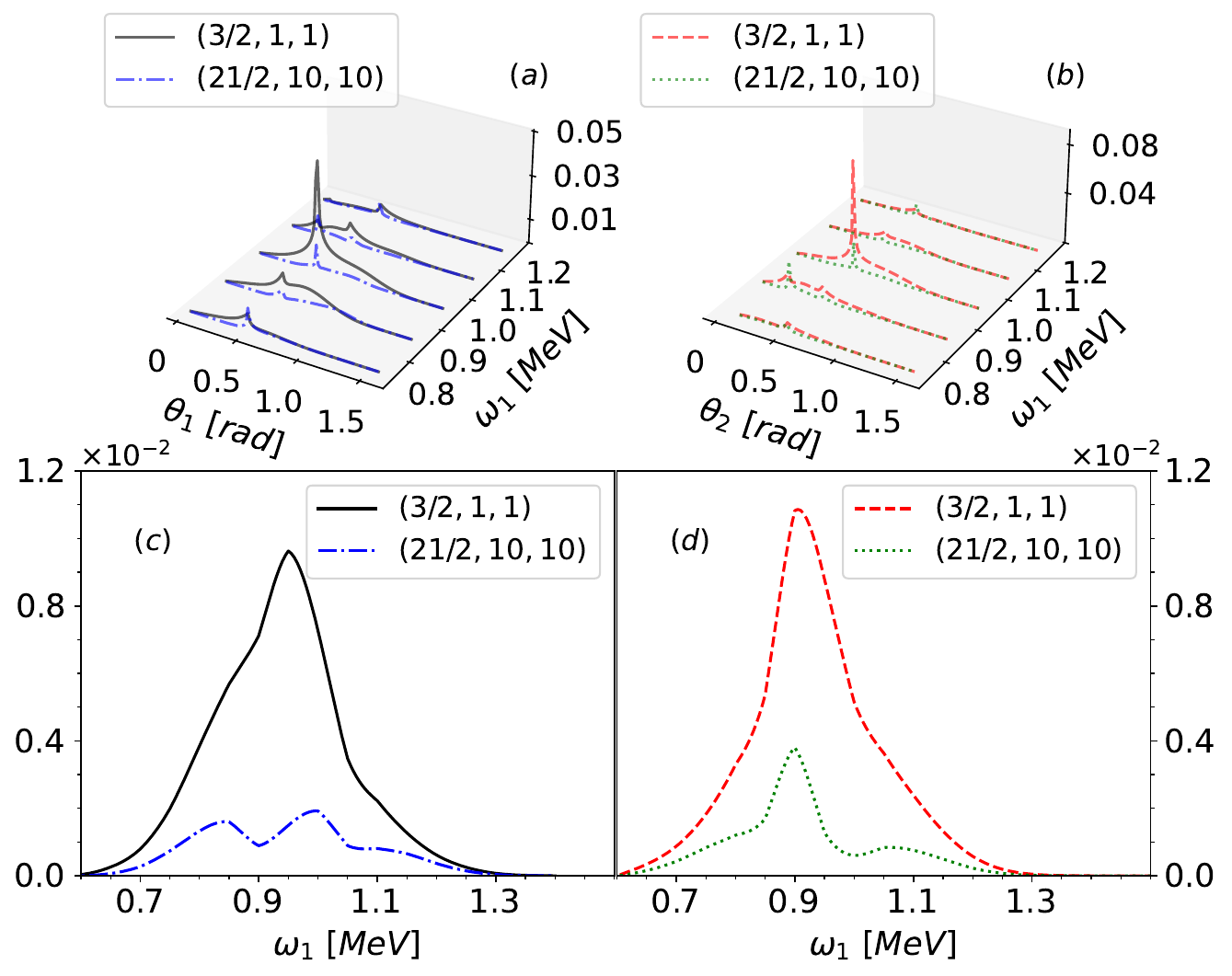}
    \caption{(a) Distribution $G_1(\omega_1;\theta_1)$ ($\textrm{b}/(\textrm{MeV }\textrm{rad}^2)$) in $\theta_1$ is shown at various values of $\omega_1$ and $(m,m_1,m_2)=(3/2,1,1),~(21/2,10,10)$. 
    (b) Similar to (a) but for $G_2(\omega_1;\theta_2)$. 
    (c) Distribution $H_1(\omega_1)$ ($\textrm{b}/(\textrm{MeV rad})$) in $\omega_1$. 
    (d) Similar to (c) but for $H_2(\omega_1)$. 
    In all panels $\theta=\theta_0=\pi/6$ and $\alpha=2\pi/3$.}
    \label{fig_alpha}
\end{figure}

So far, our discussion only concerns the case in which the final two 
photons share the same cone angle $\theta_{1}=\theta_{2}=\theta_{\gamma}$. In the following, we study the case when one of the cone angles is fixed at $\theta_0$. In order to do this, we define
\begin{subequations}
\begin{eqnarray}
G_a(\omega_1;\theta_a)&=&
G(\omega_1;\theta_1,\theta_2)\big|_{\theta_{\bar a}=\theta_0},
\\
H_a(\omega_1)&=&\int d\theta_a~G_a(\omega_1;\theta_a),
\end{eqnarray}
\end{subequations}
where $\bar a=2,~1$ for $a=1,~2$, and the results for $G_a(\omega_1;\theta_a)$ and $H_a(\omega_1)$ are presented in \cref{fig_angle,fig_angle10,fig_alpha}. As shown in \cref{fig_angle,fig_angle10}, the shape of the distribution $G_a(\omega_1;\theta_a)$ depends strongly on the fixed value of $\theta_{\bar a}=\theta_0$. And it can also be observed from the panels (a,b) in \cref{fig_angle,fig_angle10}, that the region 
in which $\theta_1$ is close to $\theta_2$ makes the major contribution to the cross section. Concerning the energy distribution $H_a(\omega_1)$, it can be seen from the panels (c,d) in \cref{fig_angle,fig_angle10} that the range of the photon energy which has a significant contribution increases with $\theta_0$ and that the multi-peak structure becomes more pronounced when increasing $\theta_0$. Comparisons of the distributions $G_a(\omega_1;\theta_a)$ between the scattering angle $\alpha=2\pi/3$ (panels (a,b) in \cref{fig_alpha}) and $\alpha=\pi/6$ (panels (a,b) in \cref{fig_angle,fig_angle10}) show that the dominant cone angles of the vortex photons are not much affected by the scattering angle $\alpha$. At $\alpha=2\pi/3$ the region in which $\theta_1$ is similar to $\theta_2$ still contributes most to the cross section, although the details of the distributions $G_a(\omega_1;\theta_a)$ are different from the ones at $\alpha=\pi/6$. However, comparisons of the energy distributions $H_a(\omega_1)$ between the two scattering angles in panels (c,d) of \cref{fig_angle,fig_angle10,fig_alpha} demonstrate a strong effect of the scattering angle $\alpha$ on the photon energy distribution. The photon energy distribution becomes broader with the increase of $\alpha$. Furthermore, the shape of 
$H_a(\omega_1)$ at $\alpha=2\pi/3$ with a given fixed $\theta_{\bar a}$ is similar to that at $\alpha=\pi/6$ with a larger fixed $\theta_{\bar a}$. These features clearly show the strong impact of the scattering angle on the process.

\section{Conclusion}
\label{sec:conclusion}

We have investigated for the first time off-axis multiple vortex scattering for a physical process, i.e. the triple-vortex process $e^-e^+\to\gamma\gamma$, in which the positron is in a plane-wave state and the electron and photons are in a vortex state. We have developed a theoretical formalism, which allows us to study the effects of various parameters such as scattering angle and orbital helicities and cone angles of vortex states. A Bessel-Gaussian type wave packet has been adopted to describe the initial vortex electron for the purpose of physical normalization while a Bessel vortex is assumed for the photons for simplicity. We have performed numerical calculations for an electron with a moderate energy of $1~\textrm{MeV}$ which head-on annihilates a positron to produce a pair of photons that are off-axis with respect to the collision axis. Our results have verified that off-axis vortex scattering is not restricted by orbital helicity conservation, in contrast to on-axis vortex scattering that has been extensively studied in the literature. Our numerical results have demonstrated strong impacts of the scattering angle, the topological charges of the initial and final vortex particles on various differential cross sections in the cone angles and energies of the final vortex photons, and thus have provided insights into the off-axis vortex scattering. The characteristic patterns observed in differential cross sections could provide a handle to distinguish or detect vortex particles.

\section*{Acknowledgements}

We thank Hao-Lin Wang for help with running {\it Mathematica} and drawing \cref{fig:HLW}. This work was supported in part by Grants  
No.\,NSFC-12035008 and No.\,NSFC-12475122.

\appendix

\section{Normalization of Bessel vortex electron and initial flux}
\label{sec:appendix_A}

The probability density for a wavefunction $\psi(\rvec,t)$ is defined as usual, $\rho(\rvec,t)=\psi^{\dagger}(\rvec,t)\psi(\rvec,t)$. In the low-spread approximation of momentum in a wave packet, the initial flux is defined as~\cite{Kotkin:1992bj,Liu:2022nfq,Karlovets:2016jrd}:
\begin{equation}
L=|v_- -v_+|\int \rho_-(\rvec,t)\rho_+(\rvec,t)~d^{3}\rvec dt,
\end{equation}
where the subscripts $\mp$ refer to the electron and positron. For a plane-wave positron normalized to one particle in the whole space, one has $\rho_+(\rvec,t)=1/V$ where $V$ is the space volume. For a Bessel-vortex electron in \eqref{eq:Bessel_state} normalized to one particle in a large cylinder of radius $R$ and length $L_z$, $\rho_-(\rvec,t)$ depends only on the radius $r$ in the transverse plane in the form of squared Bessel functions of the first kind. Using the relation 
\begin{equation}
\label{eq:Bessel_transverse_integral}
\int_0^R rdr\big(J_\ell(\kappa r)\big)^2=\frac{R}{\pi\kappa},~~~R\to\infty,
\end{equation}
the normalization constant is found to be
\begin{equation}
N_{e,\vx}=\frac{1}{\sqrt{2E_k}}\sqrt{\frac{\pi}{RL_{z}}},
\end{equation}
and the initial flux is trivially proportional to it
\begin{equation}
L=|v_- -v_+|\frac{T}{V},
\end{equation}
where $T$ is the span of time and $|v_- -v_+|$ is the magnitude of the relative velocity between the electron and positron propagating in the same axis.

\section{Amplitude for general plane-wave scattering}
\label{sec:appendix_B}

In this section, we record the explicit results of the scattering amplitude for the plane-wave process: 
\begin{eqnarray}
e^-(k,h_-) + e^+(p,h_+) \to \gamma(k_1,\lambda_1) + \gamma(k_2,\lambda_2),
\end{eqnarray}
where the electron and positron have momenta $p=(E_p,\pvec),~k=(E_k,\kvec)$ and helicities $h_\mp/2$, and the photons have momenta $k_{1,2}=(\omega_{1,2},\kvec_{1,2})$ and helicities $\lambda_{1,2}$. The tree-level amplitude in QED is, 
\begin{eqnarray}
\calA_\pw&=&\frac{e^2}{2\hat{k}_2\cdot p}
\big(B +2\tau_2\hat\pvec\cdot\vec\epsilon^*_2 A\big)
+(k_1\leftrightarrow k_2, \lambda_1\leftrightarrow\lambda_2),
\end{eqnarray}
where 
\begin{subequations}
\begin{eqnarray}
&&A=\vbar(p,h_+)\slashed{\epsilon}^*_1u(k,h_-),
\\
&&B=\vbar(p,h_+)\slashed{\epsilon}^*_2
\slashed{\hat{k}}_2\slashed{\epsilon}^*_1u(k,h_-),
\\
&&\hat k_2=\omega_2^{-1}k_2=(1,\hat\kvec_2),~~~\tau_2=\frac{|\pvec|}{\omega_2}.
\end{eqnarray}
\end{subequations}

To facilitate calculation for vortex scattering, we have to compute the above amplitude for general orientations of the momenta. As shown in \cref{fig:HLW} for the vortex scattering geometry, the positron moves in the $-z$ direction and the electron has polar and azimuthal angles $(\theta,\varphi)$ in the initial-state $xyz$ frame. The final-state $x'y'z'$ frame is specified by requiring that the two vortex photons move back-to-back in the $z'$ direction and that the $y'$ axis coincide with the $y$ axis. The two plane-wave photons have  polar and azimuthal angles $(\theta_a,\varphi_a)$ ($a=1,~2$) in the $x'y'z'$ frame. 

The bispinors of the initial electron and positron in the Dirac representation are, 
\begin{subequations}
\begin{eqnarray}
&&u(k,h_-)=\left(\begin{array}{r}
\epsilon^k_+ w(k,h_-) \\
h_-\epsilon^k_- w(k,h_-)
\end{array}\right),
\label{eq:bispinor}
\\
&&
v(p,h_+)
=\left(\begin{array}{r}
h_+\epsilon^p_-w'(p,h_+) \\
\epsilon^p_+w'(p,h_+) 
\end{array}\right),
\end{eqnarray}
\end{subequations}
where the two-component spinors are 
\begin{subequations}
\begin{eqnarray}
&&
w(k,h_-)=\frac{1}{\sqrt{2}}\left(\begin{array}{r}
h_-z\sqrt{1+h_-\cos\theta} \\ 
z^*\sqrt{1-h_-\cos\theta}
\end{array}\right),
\\
&&
w'(p,h_+)=\frac{1}{2}
\left(\begin{array}{r}
1-h_+ \\ 1+h_+
\end{array}\right),
\end{eqnarray}
\end{subequations}
with 
\begin{eqnarray}
z=e^{-i\varphi/2},\quad 
\epsilon_\pm^k=\sqrt{E_k\pm M},\quad
\epsilon_\pm^p=\sqrt{E_p\pm M}.
\end{eqnarray}
The component forms of the momenta and polarizations of the photons in the $x'y'z'$ frame are 
\begin{subequations}
\begin{eqnarray}
&&\hat\kvec'_a=(\sin\theta_a\cos\varphi_a,\sin\theta_a\sin\varphi_a,\cos\theta_a)^{\prime T},
\\
&&\vec\epsilon'_{a\lambda_a}
=\vec\eta'_{(\lambda_a)}e^{-i\lambda_a\varphi_a}c_a^2
+\vec\eta'_{(-\lambda_a)}e^{+i\lambda_a\varphi_a}s_a^2 
+\vec\eta'_{(0)}\lambda_a\sqrt{2}c_a s_a,\
\label{eq:polarization}
\\
&&\vec\eta'_{(\lambda)}=-\frac{1}{\sqrt{2}}(\lambda,i,0)^{\prime T},~
\vec\eta'_{(0)}=(0,0,1)^{\prime T},
\nonumber
\end{eqnarray}
\end{subequations}
where a prime is used to indicate components in the primed frame, with
\begin{subequations}
\begin{eqnarray}
&&c_a=\cos\frac{\theta_a}{2},\quad s_a=\sin\frac{\theta_a}{2},\quad 
c_e=\cos\frac{\theta}{2},\quad s_e=\sin\frac{\theta}{2},
\\
&&c_\alpha=\cos\alpha,\quad s_\alpha=\sin\alpha,
\end{eqnarray}
\end{subequations}
where $s_e,~c_e$ will enter through $w(k,h_-)$ using $\sqrt{1+\cos\theta}=\sqrt{2}c_e$ and $\sqrt{1-\cos\theta}=\sqrt{2}s_e$ and $c_\alpha,~s_\alpha$ will appear right now. Since $\calA_\pw$ is expressed explicitly in terms of products of three-vectors, it is easiest to rotate a component-form vector, either a momentum or a polarization vector, $\vvec'$ in the primed final-state frame to one $\vvec$ in the unprimed initial-state frame by 
\begin{eqnarray}
\vvec=R_\alpha\vvec';~~~
\vvec=\left(\begin{array}{c}
v_x \\ v_y \\ v_z
\end{array}\right),~~~
\vvec'=\left(\begin{array}{c}
v_{x'} \\ v_{y'} \\ v_{z'}
\end{array}\right),~~~
R_\alpha=\left(\begin{array}{ccc}
c_\alpha & & s_\alpha \\ & 1 &  \\ -s_\alpha & & c_\alpha 
\end{array}\right).
\end{eqnarray}

We list in the following the lengthy helicity amplitudes for $A$ and $B$ that are obtained with the help of {\it Mathematica} and {\it DeepSeek}. The expressions for the $A(h_-,h_+)$ term can be given for generic photon helicities $\lambda_a$. Denoting 
\begin{eqnarray}
\zeta_a=e^{-i\lambda_a\varphi_a},\quad
\xi_\pm=\epsilon^k_-\epsilon^p_- \pm\epsilon^k_+\epsilon^p_+,
\end{eqnarray}
the required scalar product is 
\begin{eqnarray}
\hat\pvec\cdot\vec\epsilon^*_2=-\epsilon_2^{z*}
=-\frac{\lambda_2}{\sqrt{2}}\big(
\zeta_2^* c_2^2s_\alpha
-\zeta_2 s_2^2s_\alpha 
+2c_2s_2 c_\alpha\big),
\end{eqnarray}
and the $A(h_-,h_+)$ functions are 
\begin{subequations}
\begin{eqnarray}
A(+,+)&=&\frac{1}{\sqrt{2}}\xi_+\Big\{
 c_1^{2} c_e \zeta_1^* z +  c_{\alpha} c_1^{2} c_e \lambda_1 \zeta_1^* z - 2 c_1 c_e \lambda_1 s_{\alpha} s_1 z +  c_e z \zeta_1 s_1^{2}
\nonumber
\\
&&
-  c_{\alpha} c_e \lambda_1 z \zeta_1 s_1^{2} 
+  c_1^{2} z^{*} \zeta_1^{*} \lambda_1 s_{\alpha} s_e 
+ 2 c_{\alpha} c_1 z^{*} \lambda_1 s_1 s_e 
-  z^{*} \zeta_1 \lambda_1 s_{\alpha} s_1^{2} s_e
\Big\},
\\
A(+,-)&=&\frac{1}{\sqrt{2}}\xi_-\Big\{
 c_1^2 c_e \zeta_1^* z \lambda_1 s_\alpha 
 + 2 c_\alpha c_1 c_e z \lambda_1 s_1 -
 c_e z \zeta_1 \lambda_1 s_\alpha s_1^2 +
 c_1^2 z^* \zeta_1^* s_e 
\nonumber
\\
&&
-
 c_\alpha c_1^2 z^* \zeta_1^* \lambda_1 s_e +
2 c_1 z^* \lambda_1 s_\alpha s_1 s_e +
 z^* \zeta_1 s_1^2 s_e +
 c_\alpha z^* \zeta_1 \lambda_1 s_1^2 s_e
\Big\},
\\
A(-,+)&=&\frac{1}{\sqrt{2}}\xi_-\Big\{
- c_1^2 c_e z^* \zeta_1^* \lambda_1 s_\alpha - 
2 c_\alpha c_1 c_e z^* \lambda_1 s_1 + 
 c_e z^* \zeta_1 \lambda_1 s_\alpha s_1^2 + 
 c_1^2 \zeta_1^* z s_e 
\nonumber
\\
&&
+ 
 c_\alpha c_1^2 \zeta_1^* z \lambda_1 s_e - 
2 c_1 z \lambda_1 s_\alpha s_1 s_e + 
 z \zeta_1 s_1^2 s_e - 
 c_\alpha z \zeta_1 \lambda_1 s_1^2 s_e
\Big\},
\\
A(-,-)&=&\frac{1}{\sqrt{2}}\xi_+\Big\{
- c_1^2 c_e z^* \zeta_1^* + 
 c_\alpha c_1^2 c_e z^* \zeta_1^* \lambda_1 - 
2 c_1 c_e z^* \lambda_1 s_\alpha s_1 - 
 c_e z^* \zeta_1 s_1^2 
\nonumber
\\
&&
- 
 c_\alpha c_e z^* \zeta_1 \lambda_1 s_1^2 + 
 c_1^2 \zeta_1^* z \lambda_1 s_\alpha s_e + 
2 c_\alpha c_1 z \lambda_1 s_1 s_e - 
 z \zeta_1 \lambda_1 s_\alpha s_1^2 s_e
\Big\}.
\end{eqnarray}
\end{subequations}
The $B(h_-,h_+;\lambda_1,\lambda_2)$ functions have no simple forms for generic photon helicities. Denoting 
\begin{eqnarray}
z_a=e^{-i\varphi_a},\quad
\xi_{\eta^k\eta^p}=(\epsilon^k_-+\eta^k\epsilon^k_+)(\epsilon^p_-+\eta^p\epsilon^p_+),\quad
\eta^{k,p}=\pm,
\end{eqnarray}
we list them for completely specified helicities: 
\begin{subequations}
\begin{eqnarray}
B(+,+;+,+)&=&\xi_{++}
\Big\{ 
-c_1 c_2^2 c_e z z_2^* s_1 - c_\alpha c_1 c_2^2 c_e z z_2^* s_1 + c_2^2 c_e z z_1^* z_2^* s_\alpha s_1^2 + c_1^2 c_2 c_e z z_1 s_2
\nonumber
\\
&&
+ c_\alpha c_1^2 c_2 c_e z z_1 s_2
- c_2 c_e z z_1^* s_1^2 s_2 + c_\alpha c_2 c_e z z_1^* s_1^2 s_2 - c_1^2 c_e z z_1 z_2^* s_\alpha s_2^2 + c_1 c_e z z_2^* s_1 s_2^2
\nonumber
\\
&&
- c_\alpha c_1 c_e z z_2^* s_1 s_2^2 - c_1 c_2^2 z^* z_2^* s_\alpha s_1 s_e - c_2^2 z^* z_1^* z_2^* s_1^2 s_e - c_\alpha c_2^2 z^* z_1^* z_2^* s_1^2 s_e + c_1^2 c_2 z^* z_1 s_\alpha s_2 s_e
\nonumber
\\
&&
+ 2 c_1 c_2 z^* s_1 s_2 s_e + c_2 z^* z_1^* s_\alpha s_1^2 s_2 s_e - c_1^2 z^* z_1 z_2^* s_2^2 s_e + c_\alpha c_1^2 z^* z_1 z_2^* s_2^2 s_e - c_1 z^* z_2^* s_\alpha s_1 s_2^2 s_e \Big\},
\\
B(+,+;-,+)&=&\xi_{++}
\Big\{ 
c_1^2 c_2^2 c_e z z_1^* z_2^* s_\alpha + c_1 c_2^2 c_e z z_2^* s_1 + c_\alpha c_1 c_2^2 c_e z z_2^* s_1 - c_1^2 c_2 c_e z z_1^* s_2
\nonumber
\\
&&
+ c_\alpha c_1^2 c_2 c_e z z_1^* s_2 + c_2 c_e z z_1 s_1^2 s_2 + c_\alpha c_2 c_e z z_1 s_1^2 s_2 - c_1 c_e z z_2^* s_1 s_2^2 + c_\alpha c_1 c_e z z_2^* s_1 s_2^2
\nonumber
\\
&&
- c_e z z_1 z_2^* s_\alpha s_1^2 s_2^2 - c_1^2 c_2^2 z^* z_1^* z_2^* s_e - c_\alpha c_1^2 c_2^2 z^* z_1^* z_2^* s_e + c_1 c_2^2 z^* z_2^* s_\alpha s_1 s_e + c_1^2 c_2 z^* z_1^* s_\alpha s_2 s_e
\nonumber
\\
&&
- 2 c_1 c_2 z^* s_1 s_2 s_e + c_2 z^* z_1 s_\alpha s_1^2 s_2 s_e + c_1 z^* z_2^* s_\alpha s_1 s_2^2 s_e - z^* z_1 z_2^* s_1^2 s_2^2 s_e + c_\alpha z^* z_1 z_2^* s_1^2 s_2^2 s_e 
\Big\},
\\
B(+,+;-,+)&=&\xi_{--}
\Big\{ 
c_1^2 c_2^2 c_e z z_1 z_2^* s_\alpha - c_1 c_2^2 c_e z z_2^* s_1 + c_\alpha c_1 c_2^2 c_e z z_2^* s_1 + c_1^2 c_2 c_e z z_1 s_2
\nonumber
\\
&&
+ c_\alpha c_1^2 c_2 c_e z z_1 s_2 - c_2 c_e z z_1^* s_1^2 s_2 + c_\alpha c_2 c_e z z_1^* s_1^2 s_2 + c_1 c_e z z_2 s_1 s_2^2 + c_\alpha c_1 c_e z z_2 s_1 s_2^2
\nonumber
\\
&&
- c_e z z_1^* z_2 s_\alpha s_1^2 s_2^2 + c_1^2 c_2^2 z^* z_1 z_2^* s_e - c_\alpha c_1^2 c_2^2 z^* z_1 z_2^* s_e + c_1 c_2^2 z^* z_2^* s_\alpha s_1 s_e + c_1^2 c_2 z^* z_1 s_\alpha s_2 s_e
\nonumber
\\
&&
+ 2 c_1 c_2 z^* s_1 s_2 s_e + c_2 z^* z_1^* s_\alpha s_1^2 s_2 s_e + c_1 z^* z_2 s_\alpha s_1 s_2^2 s_e + z^* z_1^* z_2 s_1^2 s_2^2 s_e + c_\alpha z^* z_1^* z_2 s_1^2 s_2^2 s_e 
\Big\},
\\
B(+,+;-,-)&=&\xi_{--}
\Big\{ 
c_1 c_2^2 c_e z z_2^* s_1 - c_\alpha c_1 c_2^2 c_e z z_2^* s_1 + c_2^2 c_e z z_1 z_2^* s_\alpha s_1^2 - c_1^2 c_2 c_e z z_1^* s_2
\nonumber
\\
&&
+ c_\alpha c_1^2 c_2 c_e z z_1^* s_2 + c_2 c_e z z_1 s_1^2 s_2 + c_\alpha c_2 c_e z z_1 s_1^2 s_2 - c_1^2 c_e z z_1^* z_2 s_\alpha s_2^2 - c_1 c_e z z_2 s_1 s_2^2
\nonumber
\\
&&
- c_\alpha c_1 c_e z z_2 s_1 s_2^2 - c_1 c_2^2 z^* z_2^* s_\alpha s_1 s_e + c_2^2 z^* z_1 z_2^* s_1^2 s_e - c_\alpha c_2^2 z^* z_1 z_2^* s_1^2 s_e + c_1^2 c_2 z^* z_1^* s_\alpha s_2 s_e
\nonumber
\\
&&
- 2 c_1 c_2 z^* s_1 s_2 s_e + c_2 z^* z_1 s_\alpha s_1^2 s_2 s_e + c_1^2 z^* z_1^* z_2 s_2^2 s_e + c_\alpha c_1^2 z^* z_1^* z_2 s_2^2 s_e - c_1 z^* z_2 s_\alpha s_1 s_2^2 s_e 
\Big\},
\\
B(-,+;+,+)&=&\xi_{-+}
\Big\{ 
c_1 c_2^2 c_e z^* z_2 s_\alpha s_1 + c_2^2 c_e z^* z_1^* z_2 s_1^2 + c_\alpha c_2^2 c_e z^* z_1^* z_2 s_1^2 - c_1^2 c_2 c_e z^* z_1 s_\alpha s_2
\nonumber
\\
&&
- 2 c_1 c_2 c_e z^* s_1 s_2 - c_2 c_e z^* z_1^* s_\alpha s_1^2 s_2 + c_1^2 c_e z^* z_1 z_2^* s_2^2 - c_\alpha c_1^2 c_e z^* z_1 z_2^* s_2^2 + c_1 c_e z^* z_2^* s_\alpha s_1 s_2^2
\nonumber
\\
&&
- c_1 c_2^2 z z_2 s_1 s_e - c_\alpha c_1 c_2^2 z z_2 s_1 s_e + c_2^2 z z_1^* z_2 s_\alpha s_1^2 s_e + c_1^2 c_2 z z_1 s_2 s_e + c_\alpha c_1^2 c_2 z z_1 s_2 s_e
\nonumber
\\
&&
- c_2 z z_1^* s_1^2 s_2 s_e + c_\alpha c_2 z z_1^* s_1^2 s_2 s_e - c_1^2 z z_1 z_2^* s_\alpha s_2^2 s_e + c_1 z z_2^* s_1 s_2^2 s_e - c_\alpha c_1 z z_2^* s_1 s_2^2 s_e 
\Big\},
\\
B(-,+;-,+)&=&\xi_{-+}
\Big\{
c_1^2 c_2^2 c_e z^* z_1^* z_2 + c_\alpha c_1^2 c_2^2 c_e z^* z_1^* z_2 - c_1 c_2^2 c_e z^* z_2 s_\alpha s_1 - c_1^2 c_2 c_e z^* z_1^* s_\alpha s_2
\nonumber
\\
&&
+ 2 c_1 c_2 c_e z^* s_1 s_2 - c_2 c_e z^* z_1 s_\alpha s_1^2 s_2 - c_1 c_e z^* z_2^* s_\alpha s_1 s_2^2 + c_e z^* z_1 z_2^* s_1^2 s_2^2 - c_\alpha c_e z^* z_1 z_2^* s_1^2 s_2^2 
\nonumber
\\
&&
+ c_1^2 c_2^2 z z_1^* z_2 s_\alpha s_e + c_1 c_2^2 z z_2 s_1 s_e + c_\alpha c_1 c_2^2 z z_2 s_1 s_e - c_1^2 c_2 z z_1^* s_2 s_e + c_\alpha c_1^2 c_2 z z_1^* s_2 s_e
\nonumber
\\
&&
+ c_2 z z_1 s_1^2 s_2 s_e + c_\alpha c_2 z z_1 s_1^2 s_2 s_e - c_1 z z_2^* s_1 s_2^2 s_e + c_\alpha c_1 z z_2^* s_1 s_2^2 s_e - z z_1 z_2^* s_\alpha s_1^2 s_2^2 s_e 
\Big\},
\\
B(-,+;+,-)&=&\xi_{+-}
\Big\{
-c_1^2 c_2^2 c_e z^* z_1 z_2^* + c_\alpha c_1^2 c_2^2 c_e z^* z_1 z_2^* - c_1 c_2^2 c_e z^* z_2^* s_\alpha s_1 - c_1^2 c_2 c_e z^* z_1 s_\alpha s_2
\nonumber
\\
&&
- 2 c_1 c_2 c_e z^* s_1 s_2 - c_2 c_e z^* z_1^* s_\alpha s_1^2 s_2 - c_1 c_e z^* z_2 s_\alpha s_1 s_2^2 - c_e z^* z_1^* z_2 s_1^2 s_2^2 - c_\alpha c_e z^* z_1^* z_2 s_1^2 s_2^2
\nonumber
\\
&&
+ c_1^2 c_2^2 z z_1 z_2^* s_\alpha s_e - c_1 c_2^2 z z_2^* s_1 s_e + c_\alpha c_1 c_2^2 z z_2^* s_1 s_e + c_1^2 c_2 z z_1 s_2 s_e + c_\alpha c_1^2 c_2 z z_1 s_2 s_e
\nonumber
\\
&&- c_2 z z_1^* s_1^2 s_2 s_e + c_\alpha c_2 z z_1^* s_1^2 s_2 s_e + c_1 z z_2 s_1 s_2^2 s_e + c_\alpha c_1 z z_2 s_1 s_2^2 s_e - z z_1^* z_2 s_\alpha s_1^2 s_2^2 s_e 
\Big\},
\\
B(-,+;-,-)&=&\xi_{+-}
\Big\{
c_1 c_2^2 c_e z^* z_2^* s_\alpha s_1 - c_2^2 c_e z^* z_1 z_2^* s_1^2 + c_\alpha c_2^2 c_e z^* z_1 z_2^* s_1^2 - c_1^2 c_2 c_e z^* z_1^* s_\alpha s_2
\nonumber
\\
&&
+ 2 c_1 c_2 c_e z^* s_1 s_2 - c_2 c_e z^* z_1 s_\alpha s_1^2 s_2 - c_1^2 c_e z^* z_1^* z_2 s_2^2 - c_\alpha c_1^2 c_e z^* z_1^* z_2 s_2^2 + c_1 c_e z^* z_2 s_\alpha s_1 s_2^2
\nonumber
\\
&&
+ c_1 c_2^2 z z_2^* s_1 s_e - c_\alpha c_1 c_2^2 z z_2^* s_1 s_e + c_2^2 z z_1 z_2^* s_\alpha s_1^2 s_e - c_1^2 c_2 z z_1^* s_2 s_e + c_\alpha c_1^2 c_2 z z_1^* s_2 s_e
\nonumber
\\
&&
+ c_2 z z_1 s_1^2 s_2 s_e + c_\alpha c_2 z z_1 s_1^2 s_2 s_e - c_1^2 z z_1^* z_2 s_\alpha s_2^2 s_e - c_1 z z_2 s_1 s_2^2 s_e - c_\alpha c_1 z z_2 s_1 s_2^2 s_e 
\Big\},
\\
B(+,-;+,+)&=&\xi_{+-}
\Big\{ 
-c_1 c_2^2 c_e z z_2 s_\alpha s_1 + c_2^2 c_e z z_1^* z_2 s_1^2 - c_\alpha c_2^2 c_e z z_1^* z_2 s_1^2 + c_1^2 c_2 c_e z z_1 s_\alpha s_2
\nonumber
\\
&&
- 2 c_1 c_2 c_e z^* s_1 s_2 + c_2 c_e z z_1^* s_\alpha s_1^2 s_2 + c_1^2 c_e z z_1 z_2^* s_2^2 + c_\alpha c_1^2 c_e z z_1 z_2^* s_2^2 - c_1 c_e z z_2^* s_\alpha s_1 s_2^2
\nonumber
\\
&&
- c_1 c_2^2 z^* z_2 s_1 s_e + c_\alpha c_1 c_2^2 z^* z_2 s_1 s_e - c_2^2 z^* z_1^* z_2 s_\alpha s_1^2 s_e + c_1^2 c_2 z^* z_1 s_2 s_e - c_\alpha c_1^2 c_2 z^* z_1 s_2 s_e
\nonumber
\\
&&
- c_2 z^* z_1^* s_1^2 s_2 s_e - c_\alpha c_2 z^* z_1^* s_1^2 s_2 s_e + c_1^2 z^* z_1 z_2^* s_\alpha s_2^2 s_e + c_1 z^* z_2^* s_1 s_2^2 s_e + c_\alpha c_1 z^* z_2^* s_1 s_2^2 s_e
\Big\},
\\
B(+,-;-,+)&=&\xi_{+-}
\Big\{ 
c_1^2 c_2^2 c_e z z_1^* z_2 - c_\alpha c_1^2 c_2^2 c_e z z_1^* z_2 + c_1 c_2^2 c_e z z_2 s_\alpha s_1 + c_1^2 c_2 c_e z z_1^* s_\alpha s_2
\nonumber
\\
&&
+ 2 c_1 c_2 c_e z^* s_1 s_2 + c_2 c_e z z_1 s_\alpha s_1^2 s_2 + c_1 c_e z z_2^* s_\alpha s_1 s_2^2 + c_e z z_1 z_2^* s_1^2 s_2^2 + c_\alpha c_e z z_1 z_2^* s_1^2 s_2^2
\nonumber
\\
&&
- c_1^2 c_2^2 z^* z_1^* z_2 s_\alpha s_e + c_1 c_2^2 z^* z_2 s_1 s_e - c_\alpha c_1 c_2^2 z^* z_2 s_1 s_e - c_1^2 c_2 z^* z_1^* s_2 s_e - c_\alpha c_1^2 c_2 z^* z_1^* s_2 s_e
\nonumber
\\
&&
+ c_2 z^* z_1 s_1^2 s_2 s_e - c_\alpha c_2 z^* z_1 s_1^2 s_2 s_e - c_1 z^* z_2^* s_1 s_2^2 s_e - c_\alpha c_1 z^* z_2^* s_1 s_2^2 s_e + z^* z_1 z_2^* s_\alpha s_1^2 s_2^2 s_e 
\Big\},
\\
B(+,-;+,-)&=&\xi_{-+}
\Big\{ 
-c_1^2 c_2^2 c_e z z_1 z_2^* - c_\alpha c_1^2 c_2^2 c_e z z_1 z_2^* + c_1 c_2^2 c_e z z_2^* s_\alpha s_1 + c_1^2 c_2 c_e z z_1 s_\alpha s_2
\nonumber
\\
&&
- 2 c_1 c_2 c_e z^* s_1 s_2 + c_2 c_e z z_1^* s_\alpha s_1^2 s_2 + c_1 c_e z z_2 s_\alpha s_1 s_2^2 - c_e z z_1^* z_2 s_1^2 s_2^2 + c_\alpha c_e z z_1^* z_2 s_1^2 s_2^2
\nonumber
\\
&&
- c_1^2 c_2^2 z^* z_1 z_2^* s_\alpha s_e - c_1 c_2^2 z^* z_2^* s_1 s_e - c_\alpha c_1 c_2^2 z^* z_2^* s_1 s_e + c_1^2 c_2 z^* z_1 s_2 s_e - c_\alpha c_1^2 c_2 z^* z_1 s_2 s_e
\nonumber
\\
&&
- c_2 z^* z_1^* s_1^2 s_2 s_e - c_\alpha c_2 z^* z_1^* s_1^2 s_2 s_e + c_1 z^* z_2 s_1 s_2^2 s_e - c_\alpha c_1 z^* z_2 s_1 s_2^2 s_e + z^* z_1^* z_2 s_\alpha s_1^2 s_2^2 s_e 
\Big\},
\\
B(+,-;-,-)&=&\xi_{-+}
\Big\{ 
-c_1 c_2^2 c_e z z_2^* s_\alpha s_1 - c_2^2 c_e z z_1 z_2^* s_1^2 - c_\alpha c_2^2 c_e z z_1 z_2^* s_1^2 + c_1^2 c_2 c_e z z_1^* s_\alpha s_2
\nonumber
\\
&&
+ 2 c_1 c_2 c_e z^* s_1 s_2 + c_2 c_e z z_1 s_\alpha s_1^2 s_2 - c_1^2 c_e z z_1^* z_2 s_2^2 + c_\alpha c_1^2 c_e z z_1^* z_2 s_2^2 - c_1 c_e z z_2 s_\alpha s_1 s_2^2
\nonumber
\\
&&
+ c_1 c_2^2 z^* z_2^* s_1 s_e + c_\alpha c_1 c_2^2 z^* z_2^* s_1 s_e - c_2^2 z^* z_1 z_2^* s_\alpha s_1^2 s_e - c_1^2 c_2 z^* z_1^* s_2 s_e - c_\alpha c_1^2 c_2 z^* z_1^* s_2 s_e
\nonumber
\\
&&
+ c_2 z^* z_1 s_1^2 s_2 s_e - c_\alpha c_2 z^* z_1 s_1^2 s_2 s_e + c_1^2 z^* z_1^* z_2 s_\alpha s_2^2 s_e - c_1 z^* z_2 s_1 s_2^2 s_e + c_\alpha c_1 z^* z_2 s_1 s_2^2 s_e 
\Big\},
\\
B(-,-;+,+)&=&\xi_{--}
\Big\{ 
c_1 c_2^2 c_e z^* z_2 s_1 - c_\alpha c_1 c_2^2 c_e z^* z_2 s_1 + c_2^2 c_e z^* z_1^* z_2 s_\alpha s_1^2 - c_1^2 c_2 c_e z^* z_1 s_2
\nonumber
\\
&&
+ c_\alpha c_1^2 c_2 c_e z^* z_1 s_2 + c_2 c_e z^* z_1^* s_1^2 s_2 + c_\alpha c_2 c_e z^* z_1^* s_1^2 s_2 - c_1^2 c_e z^* z_1 z_2^* s_\alpha s_2^2 - c_1 c_e z^* z_2^* s_1 s_2^2
\nonumber
\\
&&
- c_\alpha c_1 c_e z^* z_2^* s_1 s_2^2 - c_1 c_2^2 z z_2 s_\alpha s_1 s_e + c_2^2 z z_1^* z_2 s_1^2 s_e - c_\alpha c_2^2 z z_1^* z_2 s_1^2 s_e + c_1^2 c_2 z z_1 s_\alpha s_2 s_e
\nonumber
\\
&&
- 2 c_1 c_2 z^* s_1 s_2 s_e + c_2 z z_1^* s_\alpha s_1^2 s_2 s_e + c_1^2 z z_1 z_2^* s_2^2 s_e + c_\alpha c_1^2 z z_1 z_2^* s_2^2 s_e - c_1 z z_2^* s_\alpha s_1 s_2^2 s_e 
\Big\},
\\
B(-,-;-,+)&=&\xi_{--}
\Big\{ 
c_1^2 c_2^2 c_e z^* z_1^* z_2 s_\alpha - c_1 c_2^2 c_e z^* z_2 s_1 + c_\alpha c_1 c_2^2 c_e z^* z_2 s_1 + c_1^2 c_2 c_e z^* z_1^* s_2
\nonumber
\\
&&
+ c_\alpha c_1^2 c_2 c_e z^* z_1^* s_2 - c_2 c_e z^* z_1 s_1^2 s_2 + c_\alpha c_2 c_e z^* z_1 s_1^2 s_2 + c_1 c_e z^* z_2^* s_1 s_2^2 + c_\alpha c_1 c_e z^* z_2^* s_1 s_2^2
\nonumber
\\
&&
- c_e z^* z_1 z_2^* s_\alpha s_1^2 s_2^2 + c_1^2 c_2^2 z z_1^* z_2 s_e - c_\alpha c_1^2 c_2^2 z z_1^* z_2 s_e + c_1 c_2^2 z z_2 s_\alpha s_1 s_e + c_1^2 c_2 z z_1^* s_\alpha s_2 s_e
\nonumber
\\
&&
+ 2 c_1 c_2 z^* s_1 s_2 s_e + c_2 z z_1 s_\alpha s_1^2 s_2 s_e + c_1 z z_2^* s_\alpha s_1 s_2^2 s_e + z z_1 z_2^* s_1^2 s_2^2 s_e + c_\alpha z z_1 z_2^* s_1^2 s_2^2 s_e 
\Big\},
\\
B(-,-;+,-)&=&\xi_{++}
\Big\{ 
c_1^2 c_2^2 c_e z^* z_1 z_2^* s_\alpha + c_1 c_2^2 c_e z^* z_2^* s_1 + c_\alpha c_1 c_2^2 c_e z^* z_2^* s_1 - c_1^2 c_2 c_e z^* z_1 s_2
\nonumber
\\
&&
+ c_\alpha c_1^2 c_2 c_e z^* z_1 s_2 + c_2 c_e z^* z_1^* s_1^2 s_2 + c_\alpha c_2 c_e z^* z_1^* s_1^2 s_2 - c_1 c_e z^* z_2 s_1 s_2^2 + c_\alpha c_1 c_e z^* z_2 s_1 s_2^2
\nonumber
\\
&&
- c_e z^* z_1^* z_2 s_\alpha s_1^2 s_2^2 - c_1^2 c_2^2 z z_1 z_2^* s_e - c_\alpha c_1^2 c_2^2 z z_1 z_2^* s_e + c_1 c_2^2 z z_2^* s_\alpha s_1 s_e + c_1^2 c_2 z z_1 s_\alpha s_2 s_e
\nonumber
\\
&&
- 2 c_1 c_2 z^* s_1 s_2 s_e + c_2 z z_1^* s_\alpha s_1^2 s_2 s_e + c_1 z z_2 s_\alpha s_1 s_2^2 s_e - z z_1^* z_2 s_1^2 s_2^2 s_e + c_\alpha z z_1^* z_2 s_1^2 s_2^2 s_e 
\Big\},
\\
B(-,-;-,-)&=&\xi_{++}
\Big\{ 
-c_1 c_2^2 c_e z^* z_2^* s_1 - c_\alpha c_1 c_2^2 c_e z^* z_2^* s_1 + c_2^2 c_e z^* z_1 z_2^* s_\alpha s_1^2 + c_1^2 c_2 c_e z^* z_1^* s_2
\nonumber
\\
&&
+ c_\alpha c_1^2 c_2 c_e z^* z_1^* s_2 - c_2 c_e z^* z_1 s_1^2 s_2 + c_\alpha c_2 c_e z^* z_1 s_1^2 s_2 - c_1^2 c_e z^* z_1^* z_2 s_\alpha s_2^2 + c_1 c_e z^* z_2 s_1 s_2^2
\nonumber
\\
&&
- c_\alpha c_1 c_e z^* z_2 s_1 s_2^2 - c_1 c_2^2 z z_2^* s_\alpha s_1 s_e - c_2^2 z z_1 z_2^* s_1^2 s_e - c_\alpha c_2^2 z z_1 z_2^* s_1^2 s_e + c_1^2 c_2 z z_1^* s_\alpha s_2 s_e
\nonumber
\\
&&
+ 2 c_1 c_2 z^* s_1 s_2 s_e + c_2 z z_1 s_\alpha s_1^2 s_2 s_e - c_1^2 z z_1^* z_2 s_2^2 s_e + c_\alpha c_1^2 z z_1^* z_2 s_2^2 s_e - c_1 z z_2 s_\alpha s_1 s_2^2 s_e 
\Big\}.
\end{eqnarray}
\end{subequations}

\bibliography{ref}
\bibliographystyle{vortex}

\end{document}